\documentclass[useAMS,usenatbib]{mn2e}

\DeclareSymbolFont{cmletters}{OML}{cmm}{m}{it}
\DeclareMathSymbol{v}{\mathalpha}{cmletters}{"76}

\usepackage{amsmath}
\usepackage{amssymb}
\usepackage{epsfig}
\usepackage{graphicx}
\usepackage{ifthen}
\usepackage{latexsym}
\usepackage{rotating}
\usepackage{subfigure}
\usepackage{times,epsf}
\usepackage{txfonts}
\usepackage{varioref}
\usepackage{verbatim}
\usepackage{url}
\usepackage{color}
\usepackage[dvipsnames]{xcolor}
\usepackage[T1]{fontenc}

\newcommand{\be}{\begin{equation}}
\newcommand{\ee}{\end{equation}}
\newcommand{\bea}{\begin{eqnarray}}
\newcommand{\eea}{\end{eqnarray}}
\newcommand{\gdet}{\sqrt{-g}}

\newcommand\apj{Astrophysical Journal}
\newcommand\apjl{Astrophysical Journal Letters}

\newcommand\apjs{Astrophysical Journal Suppl. Ser.}

\newcommand\aap{Astronomy \& Astrophysics}

\newcommand\aj{Astronomical Journal}
\newcommand\nat{Nature}
\newcommand\prd{Physical Review D}
\newcommand\mnras{Monthly Notices of the Royal Astronomical Society}

\newcommand\pasj{Publications of the Astronomical Society of Japan}

\newcommand\ARAA{Ann. Rev. Astron. Asrophys.}
\newcommand\jqsrt{Journal of Quantitative Spectroscopy and Radiative Transfer}
\newcommand\araa{\ARAA}

\newcommand{\pder}[2]{\frac{\partial#1}{\partial#2}}

\newcommand{\koral}{\texttt{KORAL}}

\def\bE{\bar{E}}
\def\bR{\bar{R}}
\def\bu{\bar{u}}

\title[Radiation in relativistic conservative fluid dynamics codes]{
Semi-implicit scheme for treating radiation under M1 closure in 
general relativistic conservative fluid dynamics codes }
\author[A. Sadowski, R. Narayan, A. Tchekhovskoy \& Y. Zhu]
       {Aleksander Sadowski$^1$\footnotemark[1], 
	Ramesh Narayan$^{1}$\footnotemark[1], 
        Alexander Tchekhovskoy$^{2}$\footnotemark[1]
        and Yucong Zhu$^{1}$\thanks{E-mail: asadowski@cfa.harvard.edu (AS); 
rnarayan@cfa.harvard.edu (RN);	atchekho@princeton.edu (AT);  yzhu@cfa.harvard.edu (YZ);} \\
        $^1$ Harvard-Smithsonian Center for Astrophysics, 60 Garden
        St., Cambridge, MA 02134, USA\\
        $^2$ Princeton University, 410 Jadwin Hall, Princeton, NJ,
        08544, USA; Princeton Center for Theoretical Science Fellow }

\begin{document}

\maketitle

\label{firstpage}

\begin{abstract}
A numerical scheme is described for including radiation in
multi-dimensional general-relativistic conservative fluid dynamics
codes.  In this method, a covariant form of the M1 closure scheme is
used to close the radiation moments, and the radiative source terms
are treated semi-implicitly in order to handle both optically thin and
optically thick regimes. The scheme has been implemented in a conservative
general relativistic radiation hydrodynamics code \koral.  The
robustness of the code is demonstrated on a number of test problems,
including radiative relativistic shock tubes, static radiation
pressure supported atmosphere, shadows, beams of light in curved
spacetime, and radiative Bondi accretion.  The advantages of M1
closure relative to other approaches such as Eddington closure and
flux-limited diffusion are discussed, and its limitations are also
highlighted.
\end{abstract}


\section{Introduction}
\label{introduction}

Accretion disks are the power source behind many astrophysical
systems. They span a wide range of central object mass, size and type,
e.g., young stellar objects, neutron star binaries, black hole
binaries, gamma-ray bursts, active galactic nuclei, to name a few, and
exhibit a variety of different regimes of physics.  In many systems,
radiation is so intense that it strongly couples to the accreting gas
and dramatically alters the flow structure and dynamics. To correctly
infer the physics of such systems, we need models that properly take
into account the interaction of gas and radiation.

Proper treatment of radiation is especially important for black holes
(BHs) near the Eddington luminosity limit, $L_{\rm Edd}$.  Transient
black hole binaries (BHBs) approach near-Eddington accretion rates
near the peak of their outbursts \citep*{mccl06,remi06,done07}, while
some exceptional BHBs spend extended periods of time with $L\gtrsim
L_{\rm Edd}$ (e.g., SS433, \citealt{margon79,margon84}; GRS1915+105,
\citealt*{fb04}). Ultra-luminous X-ray sources have even larger
luminosities and may conceivably be highly super-Eddington
stellar-mass BHs \citep{wata01}, or intermediate mass BHs accreting at
close to Eddington \citep{miller04}. In either case, radiation must
play an important role. Finally, luminous active galactic nuclei,
especially those whose supermassive BHs (SMBHs) are growing rapidly in
mass, may be perfect examples of systems with closely coupled gas and
radiation \citep*{collin02}.

Since most of the radiation from an accretion disk originates in the
inner regions, general relativistic (GR) effects play an important
role in determining the emergent radiation spectrum.  Treating the
interaction of fluid dynamics and radiation quantitatively in GR
numerical codes is a daunting task that has been achieved only
recently and only for optically thick flows, as we discuss
below.  However, the emergent spectrum is established at an optical
depth of order unity, $\tau \sim 1$, which requires a proper treatment
of the optically-thick disk ($\tau\gg1$), the optically-thin
corona ($\tau\ll1$) if one is present, and the optically thick-to-thin
transition in between.
In this paper, we present a GR numerical radiation hydrodynamics
technique and a code which handles all three regimes of optical depth.

Depending on the mass accretion rate, a given accretion system can
switch between different spectral states, with different radiation
mechanisms dominating and with varying degrees of coupling between
radiation and gas.  At very low accretion rates, $L/L_{\rm Edd}\ll
10^{-2}$, e.g., Sgr A$^*$ the SMBH at our Galactic Center
\citep*{nym95}, the accretion flow advects most of the viscously
released energy into the BH rather than radiating the energy (some
energy probably also flows out in a wind).  Such flows are optically
thin, radiatively inefficient and geometrically thick, and are called
advection-dominated accretion flows (ADAFs, see \citealt{narayanyi94,
  narayanyi95}; \citealt{abramowiczetal95}). Analytical and
semi-analytical models are reasonably successful in accounting for
the main features in the spectra of ADAFs (e.g., \citealt*{yuan03}).
However, it was realized early on that numerical simulations are
necessary to understand properly the physics of ADAFs.

The low optical depth and the extreme radiative inefficiency of an
ADAF are a major advantage for simulations since they allow one to
solve the fluid dynamics separately from the radiation field\footnote{However, \cite{dibietal12} have shown that the accretion rate of Sgr A* is near the limit of the regime where radiative cooling may be important.}
A number
of sophisticated GR magnetohydrodynamic (GRMHD) codes have been
developed
\citep[e.g.,][]{devilliersetal03,gammieetal03,anninosetal05,delzannaetal07},
and ADAF-like models have been simulated using these (see
\citealt{narayan12,mtb12,tm12a} for recent work). To compute observables,
the output from such pure GRMHD codes are usually post-processed with
stand-alone radiation transfer schemes
\citep[e.g.][]{schnittmanetal06,shcherbakovetal10}. Alternatively, a
simple local cooling prescription is included directly into the code
\citep[e.g.,][]{fm09,dibietal12}; this is not difficult since the gas
is optically thin, so no radiative transfer is involved.

For higher accretion rates, $10^{-2}\lesssim L/L_{\rm Edd} \lesssim
0.3$, efficient cooling sets in, and the inner accretion disk
collapses into an optically thick geometrically thin accretion disk
\citep{narayanyi95,esin97,esin98,mm03,mccl06,done07}.  This state of
accretion is the best understood of all states, and its thermal
black-body-like spectrum is well-described by the standard thin disk
model with $\alpha$ viscosity \citep{ss73,novikovthorne73,
  frank02}. However, despite its many successes, the $\alpha$ disk
model cannot describe important microphysical aspects of the flow,
e.g., thermal and viscous stability, vertical structure of the disk,
and the role of magnetic fields.  Moreover, optically thick,
geometrically thin disks are much more difficult to simulate
numerically because of the strong coupling between gas and radiation.

Only a few time-dependent GRMHD numerical simulations of thin disks
have been computed so far, and all are based on an ad hoc (although
physically motivated) local cooling function which artificially
removes excess heat to keep the disk geometrically thin
(\citealt{shafee08,pennaetal10,nobleetal11}; see also
\citealt{rf08}).  Although this approach has produced valuable results
on the dynamics of the gas, it misses real physics involving
propagation of photons along curved geodesics, radiation pressure
support, radiative winds, etc.

Using a flux-limited diffusion approximation, small patches of
radiatively efficient thin accretion disks have been simulated using
the local shearing box approximation
\citep[][]{turner2003,kro07,blaes2007,blaes2011,hirose09a,hirose09b},
and even a few full disk simulations have been done with a
non-relativistic code using flux-limited diffusion \citep{ohsuga09,ohsuga11}.
However, to model a radiatively efficient disk self-consistently it is
necessary to handle the radiation field in both the optically thick
(disk interior) and optically thin (corona) limits.  This is difficult
with the flux-limited diffusion approximation which artificially
  enforces the flux to follow the local gradient of the radiative
  energy density.  Fueled by advances in radiation algorithms, more
advanced radiation moment closures are now becoming possible, which
allow accurate treatment of both optically thick and thin photon
fields in the ``instant light'' (nonrelativistic) approximation
\citep{hayesnorman03,gonzalesetal07,jiangetal12,davis12}.
However, as we discuss below, GR radiation transfer codes still
continue to rely on flux-limited diffusion or the Eddington
approximation.

At super-Eddington accretion rates, $L/L_{\rm Edd} \gtrsim 1$,
accretion flows again become radiatively inefficient. Here, the
optical depth is so large that the photon diffusion time from the disk
interior to the photosphere becomes longer than the accretion time. As
a result, most of the photons are advected with the gas into the BH,
leading to a ``slim accretion disk'' \citep{abra88}.  This 
important but poorly understood accretion state may be responsible for
much of the SMBH mass growth in the Universe; for instance, it might
explain the paradox of having $10^{9}M_\odot$ BHs already at
$z>6$, when the Universe was less than $1$ Gyr old \citep{barth03,willot05,fan06,willot10,mort11}. Super-Eddington accretion may also apply to
ultra-luminous X-ray sources \citep[e.g.,][]{wata01,kawashima12}.
Clearly, to understand the accretion physics in these systems,
radiation MHD models that self-consistently couple gas, radiation and
magnetic fields, are crucial.  Such models are also important for
measuring BH spins \citep{mccl11,straub11}, calculating the radiative
efficiency of super-Eddington accretion disks \citep[e.g.,][]{sad09}, and
understanding large-scale cosmological feedback by radiation-driven
outflows from AGN (see \citealt{fab12} for an observational review).  Some efforts have already been made
on attacking these important problems. In particular, using a
non-relativistic code and flux-limited diffusion, super-Eddington
accretion flows have been simulated and their spectra computed
\citep{ohsuga03,ohsuga06,ohsuga11,kawashima12}.

Until a few years ago, all radiation hydrodynamic and MHD simulations
of accretion disks were run with non-relativistic or special
relativistic codes.  However, physics around BHs must be studied in GR
because of the the strong gravity involved.  Recently, progress has
been made in implementing radiation into GR codes. \cite{farrisetal08}
developed a formalism for incorporating radiation under the Eddington
approximation in a conservative MHD code. This method has been
implemented in other codes \citep{zanottietal11,fragileetal12}, but
all these codes have problems because of the stiffness of radiative
source terms at large optical depths.  These stiff terms require
implicit treatment, which is prohibitively complicated in curved
space-time. \cite{roedigetal12}, extended the method by applying an
implicit-explicit Runge-Kutta numerical scheme. However, the authors
again relied on the Eddington approximation, which cannot handle
optically thin flows accurately. A more advanced method has been
recently described by \cite{shibatasekiguchi12}, who employ a
truncated moment formalism, similar to our method, for neutrino
transport in GR.

In the present paper, we describe a simple approach for including
radiation transport in GR codes. The method makes use of a key
simplification that is intrinsic to GR applications. Normally, in
non-relativistic codes, hydrodynamic or MHD signal speeds, which
determine the time step via the Courant condition, are much slower
than the speed of light.  Correspondingly, the time step that one uses
to evolve the fluid equations is much longer than the light-crossing
time across a cell. This is a problem when one wants to simulate
radiation hydrodynamical systems, especially in optically thin
regions, where one is faced with a large mis-match between the
characteristic speeds of the fluid and the radiation. Either one must
limit the time step to the light-crossing time, which prohibitively
increases the computational cost (as the fluid dynamics evolve on a
much slower time scale), or one must handle all the radiation terms
via an implicit method.  The latter inevitably couples neighboring
cells and makes the code very complicated. Moreover, it does not
easily generalize to curved space-time.

In contrast, a GR code is applied only in relativistic space-times
near BHs and neutron stars.  The time step in simulations is generally
set by applying the Courant condition to the smallest grid cell,
located at the innermost radius, where the flow is relativistic.
Thus, the normal time step in a pure GR hydro or GRMHD code is already
limited by the speed of light. Therefore, including radiation as an
extra relativistic fluid is fairly easy. In particular, the advection
terms in the radiation equations are no more difficult to compute than
the corresponding terms in the fluid and magnetic field evolution
equations. Nor does the time step need to be adjusted in any way.
Because of this large simplification, the advective radiation operator
can be treated via a standard explicit approach, just as one handles
the corresponding hydro and MHD terms.

Of course, the interactions between radiation and gas via emission,
absorption and scattering introduce their own time scales. These can
sometimes be very short, requiring implicit handling. However, these
interactions are local and can be handled via a {\it local} implicit
scheme. This is a great
simplification. It means that one can do implicit evolution
independently in each grid cell, without coupling to other
cells. Therefore, there are no space-time curvature effects to contend
with as in other more sophisticated multi-cell implicit schemes in GR.

In the work described here, we have implemented the above approach,
closing the radiation moment equations using the M1 closure scheme
\citep{levermore84,dubrocafeugeas99,gonzalesetal07}. M1 closure allows
a limited treatment of anisotropic radiation fields and works well in
both optically thick and thin regimes.  We have implemented our method
in a GR radiation hydrodynamics (GRRHD) code \koral.  The structure of
the paper is as follows: In Section~\ref{s.equations} we introduce the
equations, in Section~\ref{s.koral} we describe the numerical
algorithm used by \koral, in Section~\ref{s.tests} we present a set of
test problems which validate the scheme, and in
Section~\ref{s.summary} we discuss possible applications of the code.

\section{Equations}
\label{s.equations}

\subsection{Conservation laws}
\label{s.conservation}

A pure hydrodynamic flow is described by the following
conservation laws, 
\bea\label{eq.cont}
\hspace{1in}(\rho u^\mu)_{;\mu}&=&0,\\
\hspace{1in}(T^\mu_\nu)_{;\mu} &=& 0, \eea where $\rho$ is the gas
density in the comoving fluid frame, $u^\mu$ is the gas four-velocity
as measured in the ``lab frame'', and $T^\mu_\nu$ is the
hydrodynamical stress-energy tensor in this frame, 
\be\label{eq.tmunu}
T^\mu_\nu = (\rho+u+p)u^\mu u_\nu + p\delta^\mu_\nu, 
\ee 
with $u$ and $p$ representing the internal energy and pressure of the 
gas in the comoving frame.

In the case of radiation hydrodynamics, it is convenient to introduce the
radiation stress-energy tensor $R^\mu_\nu$
\citep[e.g.,][]{mihalasbook}, and to replace the second equation above
with the more general conservation law,
\be\label{eq.cons1}
(T^\mu_\nu+R^\mu_\nu)_{;\mu} = 0.  
\ee 
The radiation stress-energy tensor in an orthonormal frame
  comprises various moments of the specific intensity $I_\nu$,
  e.g., in the fluid frame it takes the following form,
\be 
\widehat R= \left[ \begin{array}{cc} \widehat E & \widehat F^i \\ \widehat F^j &
    \widehat P^{ij}
\end{array} \right],
\ee
where the fluid-frame quantities\footnote{Throughout
    the paper, ``widehats'' 
  denote quantities in the fluid frame and ``tildes'' denote quantites 
in the ZAMO frame.}
\bea
\hspace{1in}\widehat E&=& \int \widehat I_\nu {\,\rm d \nu \,d\Omega},\\
\hspace{1in}\widehat F^i&=& \int \widehat I_\nu {\,\rm d \nu \,d\Omega \,N^i},\\
\hspace{1in}\widehat P^{ij}&=& \int \widehat I_\nu {\,\rm d \nu \,d\Omega \,N^i\,N^j}
\eea
are the radiation energy density, the radiation flux and the radiation
pressure tensor, respectively, and $N^i$ is a unit vector in direction
$x^i$.

The fluid frame radiation stress tensor $\widehat R$ is related to the
tensor $\tilde R$ defined in the locally flat non-rotating frame, or
the zero-angular momentum frame (ZAMO) \citep{bardeen72}, by
\be\label{eq.convzamo1}
\tilde R^{\mu\nu}=\Lambda^\mu_\alpha(\tilde u)\Lambda^\nu_\beta(\tilde u)\widehat R^{\alpha\beta},
\ee
where $\Lambda$ is the Lorentz boost, 
\be
\Lambda(\tilde u)=
\left[ \begin{array}{cc}
\gamma & \gamma\tilde v^i \\
 \gamma\tilde v^j & \delta^{ij} + \frac{\tilde v^i\tilde
   v^j(\gamma-1)}{\tilde v^2}\\
\end{array} \right],
\ee
$\gamma=\tilde u^t$, $\tilde v^i=\tilde u^i/\tilde u^t$, and $\tilde
u^\mu$ is the four-velocity of the gas as measured by the locally
non-rotating observer.  

Quantities in the ZAMO frame (denoted with tildes) are related to
those in the lab frame by tensors created from the components of the
corresponding tetrads of the ZAMO, $e^\mu_{\nu}$ and $\tilde
e^{\mu}_{\nu}$, defined in \cite{bardeen72}.  The radiation tensor
transforms as
\bea\label{eq.convzamo2}
\hspace{1in}R^{\mu\nu}&=&e^\mu_{\alpha}e^\nu_{\beta}\tilde R^{\alpha\beta},\\
\hspace{1in}\tilde R^{\mu\nu}&=&\tilde e^{\mu}_\alpha \tilde e^{\nu}_\beta R^{\alpha\beta},\label{eq.rzamo}
\eea
while four-vectors transform as
\bea
\hspace{1in}u^{\mu}&=&e^\mu_{\alpha}\tilde u^{\alpha},\\
\hspace{1in}\tilde u^{\mu}&=&\tilde e^{\mu}_\alpha u^{\alpha}.
\eea


The conservation law (\ref{eq.cons1}) may be rewritten with the help
of the radiation four-force density $G^\nu$ as
\bea\label{eq.cons2}
\hspace{1in}(T^\mu_\nu)_{;\mu}&=&G_\nu,\\\nonumber
\hspace{1in}(R^\mu_\nu)_{;\mu}&=&-G_\nu,
\eea
where $G^\nu$ is given by \citep{mihalasbook},
\be
G^\nu=\int(\chi_\nu I_\nu - \eta_\nu){\,\rm d \nu \,d\Omega \,N^i},
\ee
which takes a particularly simple form in the fluid frame,
\be\label{eq.Gff}
\widehat G=
\left[ \begin{array}{c}
 \kappa (\widehat E-4\pi \widehat B)\\
 \chi \widehat F^i 
\end{array} \right].
\ee
Here, $\widehat B=\sigma T^4/\pi$ is the integrated Planck function corresponding to
the gas temperature $T$, $\sigma$ is the Stefan-Boltzmann constant,
$\chi_\nu$ and $\eta_\nu$ denote the frequency-dependent
opacity and emissivity coefficients, respectively, while $\kappa$ and
$\chi$ are the frequency integrated absorption and total opacity
coefficients, respectively. In Section~\ref{s.tests}, we occasionally 
refer to the scattering
opacity $\kappa_{\rm es}$, which is related to $\kappa$ and $\chi$ by 
\be
\chi \equiv \kappa + \kappa_{\rm es}.
\label{eq:kappaes}
\ee
The four-force $G^\mu$ may be transformed between
frames as described above, e.g., 
\be
G^{\mu}=e^\mu_{\alpha}\Lambda^{\alpha}_\gamma(\tilde u)\widehat
G^\gamma.  
\ee

The rest mass conservation equation~(\ref{eq.cont}) and the
energy-momentum conservation equations~(\ref{eq.cons2}) 
may be written in a coordinate basis in the 
following conservative form \citep{gammieetal03},
\bea\label{eq.cons3_1}
\hspace{.3in}\partial_t(\gdet\rho u^t)+\partial_i(\gdet\rho u^i)&=&0,\\\label{eq.cons3_2}
\hspace{.3in}\partial_t(\gdet T^t_\nu)+\partial_i(\gdet T^i_\nu)&=&\gdet T^\kappa_\lambda \Gamma^\lambda_{\nu\kappa} + \gdet G_\nu,\\\label{eq.cons3_3}
\hspace{.3in}\partial_t(\gdet R^t_\nu)+\partial_i(\gdet R^i_\nu)&=&\gdet R^\kappa_\lambda \Gamma^\lambda_{\nu\kappa} - \gdet G_\nu.
\eea
This formulation has a drawback for numerical computations: the terms
involving Christoffel symbols on the right, when calculated at cell
centers, will not balance the corresponding spatial derivatives on the
left (approximated under a given reconstruction scheme). This is true
even for particularly simple situations such as constant gas or
radiation pressure, and can lead to catastrophic secular errors. To
solve this issue we can either modify the values of the Christoffel symbols
suitably (Appendix A of McKinney et al. 2012) or we can reformulate the equations so as to avoid the problem. We choose
the second approach and make use of the following equations,
\bea
&&\hspace{0in}\partial_t(\rho u^t)+\partial_i(\rho u^i)=-\frac{\rho u^i}{\gdet}\partial_i (\gdet),\label{eq.hdlab1}\\
&&\hspace{0in}\partial_t(T^t_\nu)+\partial_i(T^i_\nu)=T^\kappa_\lambda
\Gamma^\lambda_{\nu\kappa} -\frac{T^i_\nu}{\gdet}\partial_i (\gdet)+ G_\nu,\label{eq.hdlab2}\\
&&\hspace{0in}\partial_t (R^t_\nu)+\partial_i(R^i_\nu)=R^\kappa_\lambda \Gamma^\lambda_{\nu\kappa} -\frac{R^i_\nu}{\gdet}\partial_i (\gdet) - G_\nu, \label{eq.radlab1}
\eea
where we assumed that the metric is static and moved its determinant out of the derivatives on
the left. In this formulation, the two terms that are expected to
cancel each other both appear as source terms and therefore are
calculated at the same location\footnote{In principle, it is
  sufficient to move the determinant out of the spatial
  derivatives in $r-$ and
  $\theta-$ components of equations~(\ref{eq.cons3_2}) and
  (\ref{eq.cons3_3}). }.

\subsection{Closure scheme}
\label{s.closure}
To close the above set of equations we need a prescription to compute
the second moments of the angular radiation intensity distribution.
Specifically, we need a prescription to write down the full radiation
stress tensor $R^{\mu\nu}$ knowing only the radiative energy density
$R^{tt}$ and the fluxes $R^{ti}$.

The simplest approach, which corresponds to assuming a nearly
isotropic radiation field, is the Eddington approximation, which in the fluid frame gives
\be\label{eq.eddapr} \widehat P^{ij}=\frac13\widehat E \delta^{ij}.  \ee 
However, the assumption of isotropic
specific intensity is good only in optically thick media. In
many astrophysical applications we are interested in radiation that
escapes from the photosphere to infinity, for which we need a
better closure scheme.

Following \cite{levermore84}, we assume that the radiation
tensor is isotropic and satisfies the Eddington closure, not in the
fluid frame, but in the orthonormal ``rest frame'' of the
radiation. The latter is defined as the frame in which the radiative
flux vanishes.  Thus, in this frame, we assume that $\bR^{tt}=\bE$,
$\bR^{ii}=\bE/3$, and all other components of $\bR$
are zero.  This leads to the M1 closure scheme.

In the radiation rest frame, the radiation stress tensor can be written 
in a compact form as
\begin{equation}
\bR^{\mu\nu} = \frac{4}{3}\bE\, \bu^\mu_R \bu^\nu_R + \frac{1}{3}\bE\,
g^{\mu\nu},
\label{eq:R}
\end{equation}
where $\bu^\mu_R = \{1,0,0,0\}$ and
$g^{\mu\nu}$ is the contravariant metric tensor, which in this frame
is given by the flat space Minkowski
metric. Since
equation~(\ref{eq:R}) is in a covariant form, it is also valid in the 
lab frame (as
well as all other frames), with
$u^\mu_R$ being the four-velocity of the radiation rest frame as
measured by an observer in the lab frame.
Note that, regardless of which frame one works in, 
the quantity $\bE$ should be interpreted as 
the radiation energy density as measured in the
\emph{radiation rest frame}.

Each time step in the numerical integration in any particular cell
gives an update to the ``time'' row of the radiation tensor,
$R^{t\nu}$, for that cell in the lab frame. Thus, we obtain numerical
values of these four particular components of the tensor.  According
to equation~(\ref{eq:R}), the full tensor $R^{\mu\nu}$ is a function
of 5 numbers, $\bE$, $u^\mu_R$, though only four of these are independent since
the norm of the four-velocity $u^\mu_R$ is equal to $-1$. Hence, we can use the
four given tensor elements to solve for the four unknowns and thereby
compute the full matrix. Below we give an algorithm for doing this
analytically.

Consider the quantity $R^{t\mu}R^{t\nu}$ which can be expressed as (equation~\ref{eq:R}),
\begin{equation}
R^{t\mu}R^{t\nu} = \frac{1}{9}\bE^2\left[16 (u^t_R)^2u^\mu_R u^\nu_R
+4 u^t_Ru^\mu_R g^{t\nu}+4 u^t_Ru^\nu_R g^{t\mu} + g^{t\mu}g^{t\nu}\right].
\end{equation}
If we contract this with $g_{\mu\nu}$ and use the following results,
\begin{eqnarray}
\hspace{2cm}g_{\mu\nu}\,u^\mu_R u^\nu_R &=& -1, \\
\hspace{2cm}g_{\mu\nu}\,u^\mu_R g^{t\nu} &=& u_{\nu,R}\, g^{t\nu} = u^t_R, \\
\hspace{2cm}g_{\mu\nu}\,g^{t\mu}g^{t\nu} &=& \delta^t_\nu\, g^{t\nu} = g^{tt},
\end{eqnarray}
we obtain
\begin{equation}
g_{\mu\nu}\,R^{t\mu}R^{t\nu} = -\frac{8}{9}\bE^2 (u^t_R)^2
+\frac{1}{9}\bE^2 g^{tt}.
\label{eq:invert1}
\end{equation}
The left-hand side is computable from the four given tensor elements
and the right-hand side involves two of the unknowns: $\bE$, $u^t_R$.
We also have the following expression for $R^{tt}$,
\begin{equation}
R^{tt} = \frac{4}{3}\bE (u^t_R)^2 + \frac{1}{3}\bE g^{tt},
\label{eq:invert2}
\end{equation}
which again involves the same two unknowns.  Thus, we can solve
equations (\ref{eq:invert1}) and (\ref{eq:invert2}) to obtain $\bE$
and $u^t_R$ (it reduces to a quadratic equation).  It is then
straightforward to calculate the remaining $u^i_R$ from the other
time components of equation~(\ref{eq:R}) and to calculate the entire
radiation stress tensor.

For flat spacetime, the above formulation reduces to the 
standard formulae \citep{levermore84,dubrocafeugeas99,gonzalesetal07}.
For instance, the radiation pressure tensor $\widehat P^{ij}$ 
in the fluid frame has the form,
\be
\widehat P^{ij}=\left(\frac{1-\xi}{2} \delta^{ij}+\frac{3\xi-1}{2} \frac{f^if^j}{|f|^2}\right)\widehat E,
\ee
where $f^i=\widehat F^i/\widehat E$ is the reduced radiative flux and $\xi$ is the
Eddington factor given by
\citep{levermore84},
\be
\xi=\frac{3+4f^if_i}{5+2\sqrt{4-3f^if_i}}.
\ee

In the extreme ``optically thick limit'', $\widehat F^i \approx 0$, 
and we find $f^i =
0$, $f^if_i = 0$ and $\xi = 1/3$, which corresponds to the correct answer,
viz., the Eddington approximation,
\be
\widehat P^{ij}_{\tau \gg 1}=
\left[ \begin{array}{ccc}
1/3&0&0\\
0&1/3&0\\
0&0&1/3\\
\end{array} \right]\widehat E.
\ee
In the opposite extreme ``optically thin limit'', $\widehat F^1=\widehat E$, 
i.e., a uni-directional radiation field directed along the x-axis, we
have $f^i = \delta^i_1$, $f^if_i = 1$ and $\xi = 1/3$, which gives
\be
\widehat P^{ij}_{\tau \ll 1}=
\left[ \begin{array}{ccc}
1&0&0\\
0&0&0\\
0&0&0\\
\end{array} \right]\widehat E,
\ee
This corresponds to an intensity distribution in the form of a
Dirac $\delta$-function parallel to the flux vector, which is
again the correct answer.  The M1 closure scheme thus handles both
optical depth extremes well and it is found to be fairly good at
intermediate optical depths as well.

As explained, the M1 closure scheme assumes that radiation is isotropic in the
radiation ``rest frame''. The stress tensor in an arbitrary frame is
obtained by applying a Lorentz boost to the isotropic ``rest frame'' tensor. As
a result, only one direction, the direction of the boost, is
distinguished. In other words, the specific intensity is always
symmetric with respect to the mean flux. The M1 closure scheme is thus
expected to be only approximate when multiple sources of light are involved (see Section~\ref{s.shadow}).
In problems involving accretion disks, which are the primary area of
interest of the present authors, highly anisotropic configurations
with multiple beams are not very common, and the M1 scheme is probably
adequate. In any case, M1 closure will provide a significantly
superior treatment of radiation in the optically thin regions near and
above the disk photosphere, compared to the Eddington approximation or
flux-limited diffusion.

\section{The \koral\ code}
\label{s.koral}
The scheme described in this paper has been implemented into a GRRHD
code \koral\ which solves equations~(\ref{eq.hdlab1})--(\ref{eq.radlab1}) in an
arbitrary metric. The code uses a finite difference scheme with either
linear slope-limited reconstruction \citep{kurganovtadmor00} or
fifth-order non-linear monotonizing filter reconstruction \citep[MP5,
  see][]{sureshhuynh97,delzannaetal07}. The fluxes at the cell faces
are calculated using the Lax-Friedrichs scheme. The source terms are
applied at the cell centers and the time stepping is performed using
the optimal Runge-Kutta method of third order \citep{shuosher88}. The
vector of conserved quantities is (Section~\ref{s.implementation})
\be U = [\rho u^t,T^t_t+\rho u^t,T^t_i,R^t_t,R^t_i], \ee while the
primitive quantities are, \be P = [\rho, u, u^i/u^t, \widehat E, \widehat F^i].  \ee
Conversion from conserved to primitive quantities is described in
Section~\ref{s.conversions}, while the algorithm itself is described
in the next one.

\subsection{The algorithm}
\label{s.algorithm}
During each sub-step of the Runge-Kutta time integration, the code
carries out the following steps in the given order:
\begin{enumerate}
\item The vector of conserved quantities in each cell is used to
  calculate the primitive quantities at the cell center (see
  Section~\ref{s.conversions}).
\item Ghost cells at the boundaries of the computational domain are
  assigned primitives appropriate to the boundary conditions of the
  particular problem of interest.
\item For each cell, the maximal characteristic left- and right-going
  wave speeds are calculated, following the algorithm described in
  Section~\ref{s.wavespeeds}.
\item For each dimension, primitives are interpolated using the chosen
  reconstruction scheme (linear slope-limiter or non-linear
  monotonizing filter) to obtain their left- and right-biased values
  at cell faces: $P_L$ and $P_R$.
\item From $P_L$ and $P_R$, left- and right-biased fluxes ${\cal F}_L$
  and ${\cal F}_R$ are calculated at cell faces.
\item The flux at a given cell face is calculated using the
  Lax-Friedrichs formula,
\be\label{eq.cons4}
{\cal F}=\frac12({\cal F}_R+{\cal F}_L-a(U_R-U_L)),
\ee
where $a$ is the maximal absolute value of the characteristic speeds
at the centers of the two cells on either side of the face, and $U_R$
and $U_L$ are the conserved quantities calculated at the cell face
based on $P_R$ and $P_L$.
\item The advective time derivative is calculated using an unsplit
  scheme,
\be\label{eq.lax}
\frac {dU}{dt}_{\rm (adv)} = -\frac{{\cal F}^1_R-{\cal F}^1_L}{dx^1}-\frac{{\cal F}^2_R-{\cal F}^2_L}{dx^2}-\frac{{\cal F}^3_R-{\cal F}^3_L}{dx^3},
\ee
where $dx^i$ denotes cell size in the direction $i$. Note that all
primitives, including the radiation density $\widehat{E}$ and radiation flux
$\widehat{F}^i$, are treated identically as far as the advective term is
concerned.
\item The geometrical source terms, viz., all terms on the right hand
  sides of equations~(\ref{eq.hdlab1})--(\ref{eq.radlab1}) except the radiation
  four-force density $\pm G_\nu$, are calculated at cell centers to
  give the corresponding time derivative ${dU}/{dt_{\rm (geo)}}$.
\item The advective and geometrical operators are used to update the
  conserved quantities according to
\be
\Delta U = \left(\frac {dU}{dt}_{\rm (adv)}  + \frac {dU}{dt}_{\rm (geo)}  \right) \Delta t.
\ee
That is, all these terms are treated in an explicit fashion.
\item The updated vectors of conserved quantities are used to
  calculate the corresponding updated primitive quantities at cell
  centers.
\item Finally, the remaining terms, viz., those involving the
  four-force density $G^\mu$, are handled implicitly using the method
  described in Section~\ref{s.radsource}.  This results in a final
  update of the vector of conserved quantities at each cell center.
\end{enumerate}

\subsection{Characteristic wavespeeds}
\label{s.wavespeeds}

The Lax-Friedrichs scheme
requires knowledge of the maximal characteristic wave speeds of the
system ($a$ in equation~\ref{eq.cons4}). The hydrodynamical and the
radiative components of equations~(\ref{eq.cons3_1})--(\ref{eq.cons3_3}) are
coupled only through the radiative source term $\pm G_\nu$. Therefore,
for the purpose of calculating the fluxes at cell faces and the
advective time derivative, we are allowed to separate the
hydrodynamical and radiative wave speeds. We can calculate each
separately, and use it for evaluating its corresponding flux. 
Such an approach avoids excessive artificial numerical viscosity which appears when the characteristic wavespeeds are not separated. The radiative wavespeed, as described below, never drops below $c/\sqrt{3}$. If such a high value was used in equation~(\ref{eq.cons4}) for the hydrodynamical subsystem, it would result in strong artificial diffusion of the gas.

The hydrodynamical characteristic velocity in the fluid frame is the
local speed of sound, \be c_s=\sqrt{\frac{\Gamma p}{\rho+u+p}}.  \ee
To get the left- and right-going wave speeds in the lab frame we
transform this velocity in the standard way (e.g.,
\citealt{gammieetal03}).

The evolution of the radiation field is described in the fluid frame
by the following set of equations,
\bea\label{eq.radff}
\hspace{.85in}\partial_t \widehat E +\partial_i\widehat  F^i &=&-\widehat G^t,\\\nonumber
\hspace{.85in}\partial_t \widehat F^j+\partial_i \widehat P^{ij}&=&-\widehat G^j, \eea
which is the non-relativistic limit of equations~(\ref{eq.cons3_3}) in flat
spacetime. Given a closure scheme, it is possible to calculate the
Jacobi matrices, \be\label{eq.J} J^i= \left[ \begin{array}{cc}
    \partial{\widehat F^i}/\partial{\widehat E} & \partial{\widehat F^i}/\partial{\widehat F^j}
    \\ \partial{\widehat P^{ij}}/\partial{\widehat E} & \partial{\widehat P^{ij}}/\partial{\widehat F^j}
\end{array} \right],
\ee 
whose eigenvalues give the wave speeds of interest in a given
direction $i$. The maximal left- and right- going speeds can then be
transformed to the lab frame following the same method as for fluid
velocities.

For the M1-closure scheme, the equation for the eigenvalues is
quartic and may be solved efficiently using standard
algorithms. Another approach, which will likely improve code
performance, is to precalculate tables of radiative wave speeds as a
function of the reduced flux vector components $\widehat F^i/\widehat E$
\citep{gonzalesetal07}. At present, \koral\ uses the
analytical approach.

In the limit of large optical depths, the radiative energy density,
when decoupled from gas (e.g., for $\kappa \ll 1$ but $\chi \gg 1$),
has a diffusion coefficient $D$ given by (see Section~\ref{s.diffusion})
\be\label{eq.Ddiff}
D=\frac1{3\chi}.
\ee
In this limit the distribution of radiative energy density should
remain stationary ($\partial/\partial t \rightarrow 0$).
On the other hand, the maximal eigenvalue of the
Jacobi matrices $J^i$ (equation~\ref{eq.J}) is $\pm1/\sqrt{3}$
\citep{gonzalesetal07}. If such large wave speeds are incorporated into
a numerical scheme they will result in large, unphysical, numerical
diffusion. To limit this effect, we modify the radiative
wave speeds in the fluid frame according to
\bea\label{eq.wavespeedlimit}
\hspace{.9in }a^i_R&\rightarrow &{\rm min}\left(a^i_R,\frac4{3\tau^i}\right),\\\nonumber
\hspace{.9in }a^i_L&\rightarrow &{\rm max}\left(a^i_L,-\frac4{3\tau^i}\right),
\eea
where $a^i_R$ and $a^i_L$ are the maximal right- and left-going
radiative wave speeds in the fluid frame in the direction $i$, and
$\tau^i=\chi {\rm dx^i}$ is the total optical depth of a given cell in
that direction.

The smaller the characteristic wave speed in equation~(\ref{eq.lax}), the
weaker the numerical diffusion. Thus, one may be tempted to set the wave speed to zero. However, the numerical scheme will then 
no longer satisfy the total variation diminishing (TVD) condition and the algorithm will be unstable.
Our choice
of the wave speed limiter (equation~\ref{eq.wavespeedlimit}) is motivated
by the fact that, for a diffusion equation of the form
$y_{,t}=Dy_{,xx}$, the maximum allowed time step for an explicit
numerical solver is
\be
\Delta  t =\frac{(\Delta x)^2}{4D}.
\ee
This expression, combined with equation~(\ref{eq.Ddiff}), gives
\be
\frac{\Delta x}{\Delta t}=\frac{4}{3\chi \Delta x}=\frac{4}{3\tau},
\ee
which is the limiter introduced in equation~(\ref{eq.wavespeedlimit}).

\subsection{Implicit treatment of radiative source terms}
\label{s.radsource}

It is well-known that, under some circumstances, e.g. for large
optical depths, the radiative source terms $\pm G_\nu$ in
equations~(\ref{eq.cons3_2}) and (\ref{eq.cons3_3}) become stiff, making
explicit integration practically impossible
\citep[e.g.,][]{zanottietal11}. We then need to treat these terms via
implicit time integration.  In principle, we could try to solve the
whole system of partial differential 
equations~(\ref{eq.hdlab1})--(\ref{eq.radlab1}) implicitly, as done for instance
in non-relativistic or special relativistic radiation hydrodynamics
codes \citep[e.g.,][]{krumholzetal07,jiangetal12}. However, this
approach is very difficult in GR, where the curvature of spacetime
makes the problem highly complicated and it is non-trivial to ensure
that an implicit code is conservative.  

As already explained, our approach is to split the advective
derivative operator from the radiative source terms operator. The former is
applied explicitly in the usual way while the latter is handled implicitly.
This approach is possible because the time step is
already limited by the speed of light just from the fluid dynamics,
so radiation advection is also guaranteed to be stable in an explicit
scheme.\footnote{
    Note that if we were to apply the code to non-relativistic problems, or
  situations involving weak gravity, there would be a large disparity
  in the advective time scales of fluid and radiation, and the code
  would no longer be efficient.}  The main advantage is that the
radiation source terms $\pm G^\nu$ are local, so they can be treated
semi-implicitly and point-wise in the fluid frame, without having to
deal with curvature effects.  In our experience, this approach is both
simple and robust.

The radiative source term operator describes the action of the
radiative four-fource $G^\nu$ on the energy and momentum density
of the gas and the radiation.  The corresponding
equations are
\bea
&&\hspace{2cm}\partial_t(T^t_\nu)=G_\nu,\label{eq.source1}\\
&&\hspace{2cm}\partial_t (R^t_\nu)= - G_\nu. \label{eq.source2}
\eea
In an explicit scheme, updates would be calculated very simply as
\bea
&&\hspace{1cm}T^t_{\nu,(n+1)}-T^t_{\nu,(n)}=\Delta t ~G_{\nu,(n)},\\
&&\hspace{1cm}R^t_{\nu,(n+1)}-R^t_{\nu,(n)}= - \Delta t~ G_{\nu,(n)},
\eea
where the subscripts $(n)$ and $(n+1)$ denote values at the
beginning and end of a time step of length $\Delta t$, respectively.
This approach, though simple, 
is numerically very unstable whenever the terms in
the force vector $G_\nu$ are large. Our scheme avoids
the instability by computing the updates \emph{implicitly} via
\bea
&&\hspace{1cm}T^t_{\nu,(n+1)}-T^t_{\nu,(n)}=\Delta t ~G_{\nu,(n+1)},\label{eq.source3}\\
&&\hspace{1cm}R^t_{\nu,(n+1)}-R^t_{\nu,(n)}= - \Delta t~ G_{\nu,(n+1)}, \label{eq.source4}
\eea
i.e., using quantities at time $(n+1)$ rather than $(n)$ to compute
the force vector on the right-hand side. It is well-known that this
simple change has a profound effect on stability.

\koral\ solves equations~(\ref{eq.source3}) and (\ref{eq.source4})
numerically. Because of the symmetry of the problem, specifically, the
right-hand sides of eqs.~(\ref{eq.source3})
and (\ref{eq.source4}) differing only by sign,
the system of equations may be reduced to four
non-linear equations, e.g., for $R^t_{\mu,(n+1)}$. We use the standard
Newton method to solve these equations 
and estimate the Jacobian matrix numerically. During
each iteration we use the current guess of $R^t_{\mu,(n+1)}$ to
calculate the corresponding values of $T^t_{\mu,(n+1)}$, \be
T^t_{\mu,(n+1)}=T^t_{\mu,(n)}-(R^t_{\mu,(n+1)}-R^t_{\mu,(n)}).  \ee
We then convert the current vector
of conserved quantities to primitives (see Section~\ref{s.conversions})
and calculate the radiative four force in the fluid frame,
$\widehat G^\nu_{(n+1)}$. This is then boosted to the laboratory frame
to obtain $G_{\nu,(n+1)}$.

The method described above is numerical and can occasionally fail.
In Appendix~\ref{ap.expimp} we describe an alternate, fully
analytical, but approximate, method for applying
the radiative source operator. \koral\  uses that algorithm as a
failsafe backup whenever the numerical scheme descibed in
this section fails.

\subsection{Conversion between conserved and primitive quantities}
\label{s.conversions}
In conservative GR numerical codes it is necessary to convert
conserved quantities to primitives at least once per sub-time step
(twice in the case of our algorithm). In our problem, the
hydrodynamical and radiative variables decouple, so the conversion may
be done separately for each.

The conversion of hydrodynamical quantities is performed in the usual
manner \citep[e.g.,][]{nobleetal06,delzannaetal07}. We take the
explicit forms of the conserved quantities (equation~\ref{eq.tmunu}), use
the equation of state $p=(\Gamma-1)u$, the four-velocity normalization
$u^\mu u_\mu=-1$, and combine all the terms into a non-linear equation
for the internal energy density $u$. We solve this equation
numerically by the Newton method using the value of $u$ from
the end of the previous time step as the initial guess.

The radiative conserved variables ($R^t_\mu$) may be easily converted to
the radiative primitives ($\widehat E$, $\widehat F^i$). First, one has to calculate all
the components of the radiation stress tensor in the lab frame
$R^{\mu\nu}$ following the algorithm described in Sect.~\ref{s.closure}.
This tensor is then boosted to the fluid frame (Sect.~\ref{s.conservation}),
\be \widehat R^{\mu\nu}=\Lambda^\mu_\alpha(-\tilde
u)\Lambda^\nu_\beta(-\tilde u) \tilde e^{\alpha}_\kappa \tilde
e^{\beta}_\lambda R^{\kappa\lambda}.  \ee 
The
fluid-frame radiative energy density $\widehat E$ and fluxes $\widehat
F^i$ are then
given by, \bea
\hspace{1in}\widehat E&=&\widehat R^{tt},\\
\hspace{1in}\widehat F^i&=&\widehat R^{ti}.
\eea

\subsection{Implementation notes}
\label{s.implementation}
\begin{enumerate}
\item 
The mass conservation equation~(\ref{eq.hdlab1}) and the gas internal
energy conservation law, i.e., the $t$ component of
equation~(\ref{eq.hdlab2}), are aggregated to give \citep{gammieetal03}
\be
\partial_t\left(T^t_t+ \rho u^t\right)+\partial_i\left( T^i_t+
\rho u^i\right)=T^\kappa_\lambda \Gamma^\lambda_{t\kappa} -\frac{T^i_\nu+\rho u^i}{\gdet}\partial_i (\gdet)+ G_t,\label{eq.hdlab4}
\ee
which replaces the $t$ component of equation~(\ref{eq.hdlab2}).  Then,
$(T^t_t + \rho u^t)$ becomes the relevant conserved quantity, which
reduces in the non-relativistic limit to $T^t_t + \rho u^t\rightarrow -u$.
\item In cold relativistic flows, where $u\ll \rho$, the numerical
  accuracy is not suficient to evolve the internal energy reliably.
  As a result, negative internal energy densities may be occasionally
  found. Whenever this happens, we recalculate the gas properties by
  evolving the gas entropy per unit volume, \be S=\frac{\rho}{(\Gamma-1)}
  \log\left(\frac{p}{\rho^\Gamma}\right) .  \ee That is, we compute
  the entropy at the end of the last successful time step and evolve
  it according to $(S u^\mu)_{;\mu}=0$.

\end{enumerate}

\section{Test problems}
\label{s.tests}


\subsection{Relativistic shock tube}
\label{s.hdtube}

To validate the hydrodynamics part of our code, we tested it using the
relativistic shock tube problem introduced by
\cite{hawleyetal84a}. The modeled system consists of two states (left and
right), separated initially by a membrane.  The gas on the
left is dense and hot, while that on the right is rarefied and cool
(Table~\ref{t.hdtube}). At time $t=0$ the membrane is removed and the
hot gas pushes the cool gas to the right causing a shock to travel to
the right. Meanwhile, a rarefaction wave moves at the local sound
speed back to the left. An analytical solution for this shock tube
problem may be obtained by solving the appropriate jump conditions.

\begin{figure}
  \centering
\includegraphics[width=.8\columnwidth,angle=270]{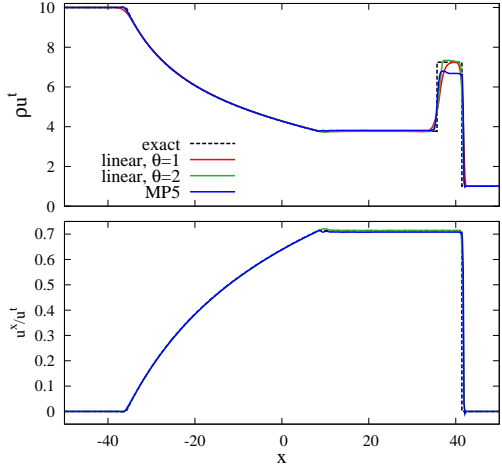}
\caption{The relativistic shock tube test solved on 500 grid points
  using three reconstruction algorithms: slope-limited linear with
  $\theta=1$ (most diffusive, red), $\theta=2$ (green), and MP5
  (blue). The black dotted line shows the exact solution for comparison.}
  \label{f.hdtube}
\end{figure}

In Figure~\ref{f.hdtube} we show our numerical solutions at time
$t=50$, which may be compared with the exact solution (black dashed
line) obtained using an exact Riemann solver
\citep{giacomazzorezzolla06}. The upper panel shows profiles of
density obtained using three different reconstruction schemes: linear
slope limiter with $\theta=1$, linear slope limiter with
$\theta=2$,\footnote{The $\theta$ parameter in the generalized minmod
  limiter determines the diffusivity of the scheme
  \citep{kurganovtadmor00}; $\theta = 2$ corresponds to the MC
  (monotonized central) scheme and $\theta = 1$ corresponds to the
  MINMOD scheme.} and fifth-order monotonizing filter, MP5. The
rarefaction wave and the plateau are resolved equally well in all
three schemes. However, there are differences in the post shock
region. The linear $\theta=1$ scheme (red lines) is most diffusive and
smooths the edges of the postshock region, while the MP5 scheme (blue)
somewhat underestimates the density here.
The lower
panel shows the velocity. This is reproduced well in all three
schemes, though the MP5 scheme produces low-level oscillations near
the edge of the rarefaction wave.

\begin{table}
\caption{Model parameters for relativistic hydrodynamical shock tube test}
\label{t.hdtube}
\centering\begin{tabular}{@{}cccccccccc}
\hline
 &  & \multicolumn{3}{c}{Left state:}&  &\multicolumn{3}{c}{Right state:} \\
 $\Gamma$ &  &$\rho$ & $p$ & $u^x$  &  &  $\rho$ & $p$ & $u^x$  \\
\hline
5/3& &10.0 & 13.33 & 0.0 & & 1.0 & $10^{-8} $&0.0 \\
\hline
\end{tabular}
 \end{table}

This test shows that all the reconstruction schemes currently
implemented in \koral\ work reasonably well on relativistic
hydrodynamic shocks. The MP5 scheme is most accurate in smooth regions, but it is also
prone in some circumstances to give unphysical oscillations. The
$\theta=1$ linear scheme is most diffusive, and at the same time is
also the most stable. In the following tests, if not stated otherwise,
we use the linear reconstruction scheme with $\theta=1.5$. 

\subsection{Polish doughnuts}

To test the ability of the code to handle multi-dimensional
hydrodynamics in curved spacetime, we set up analytical equilibrium
torii \citep[Polish doughnuts,][]{abramowiczetal78} in the
Schwarzschild metric as initial conditions and let the code evolve to
a numerical equilibrium configuration. For the analytical model, we
assume a constant specific angular momentum, $\ell =
-u_\phi/u_t=$\, constant. From the condition $u^\mu
u_\mu=-1$, it follows that \be u_t^{-2}=-g^{tt}+2\ell
g^{t\phi}-\ell^2g^{\phi\phi}.  \ee We choose the specific
internal energy at the inner edge of the torus, $u_{t,\rm in}$, which
determines the radius of the inner edge of the torus, and we then
calculate the fluid enthalpy, $h=\rho+u+p$
\citep[e.g.,][]{hawleyetal84a}, \be h=\frac{u_{t,\rm in}}{u_t}.  \ee
Using an equation of state $p=\kappa \rho^\Gamma$ (where the constant
$\kappa$ determines the entropy of the torus gas), we obtain \bea
\hspace{1in}\rho&=&\left[\frac{(h-1)}{\kappa}\,\frac{(\Gamma-1)}{\Gamma}\right]^{1/(\Gamma-1)}, \\
\hspace{1in}u&=&\rho\,\frac{(h-1)}{\Gamma}.
\eea
We set the initial velocity to $v^r=v^\theta=0$, $v^\phi=u^\phi/u^t$,
and choose $\Gamma=4/3$. 

Table~\ref{t.doughnuts} gives parameter values corresponding to three
models that we ran to test the code. Models 1 and 2 have the same
value of the specific angular momentum, $\ell=4.5$. Model 1 has
$u_{t,\rm in}=-1$, corresponding to a torus inner radius $r_{\rm
  in}=8$, while Model 2 has $u_{t,\rm in}=-0.98$, corresponding
to $r_{\rm in}=10$. The specific angular momentum for the third
model, $\ell=3.77$, lies in between the Keplerian values of $\ell$ at
the marginally stable and marginally bound orbits ($\ell_{ms} < \ell <
\ell_{mb}$). Therefore, this torus has a cusp (self-crossing
equipotential surface) near the marginally bound orbit
($r_{mb}=4$). All the equipotential surfaces outside the body of the
torus (defined by the critical surface producing the cusp) are open 
and continue into the BH. Therefore, the assumption of zero poloidal velocity cannot be
applied to this region.
\begin{table}
\caption{Model parameters for Polish doughnuts tests}
\label{t.doughnuts}
\centering\begin{tabular}{@{}lccc}
\hline
 Model & $\ell$ & $u_{\rm t, in}$ & $\kappa$ \\
\hline
1 & 4.5 & -1.0 & 0.03 \\
2 & 4.5 & -0.98& 0.03 \\
3 & 3.77& -1.0 & 0.03 \\
\hline
\end{tabular}
 \end{table}

The above three torus models were simulated on a 50x50 grid in
Boyer-Lindquist ($r$-$\theta$) coordinates, linearly spanning the
range $r=3-27.8$ and $\theta=0-\pi/2$. At the spin axis and the
equatorial plane, reflection boundary conditions were assumed. The
analytical solution was imposed as boundary conditions in the ghost
cells outside $r=27.8$. At the inner edge of the grid $r=3$, outflow
boundary conditions were implemented. Linear reconstruction with
$\theta=1.5$ was used.

\begin{figure}
  \centering
\subfigure{\includegraphics[width=.545\columnwidth,angle=270]{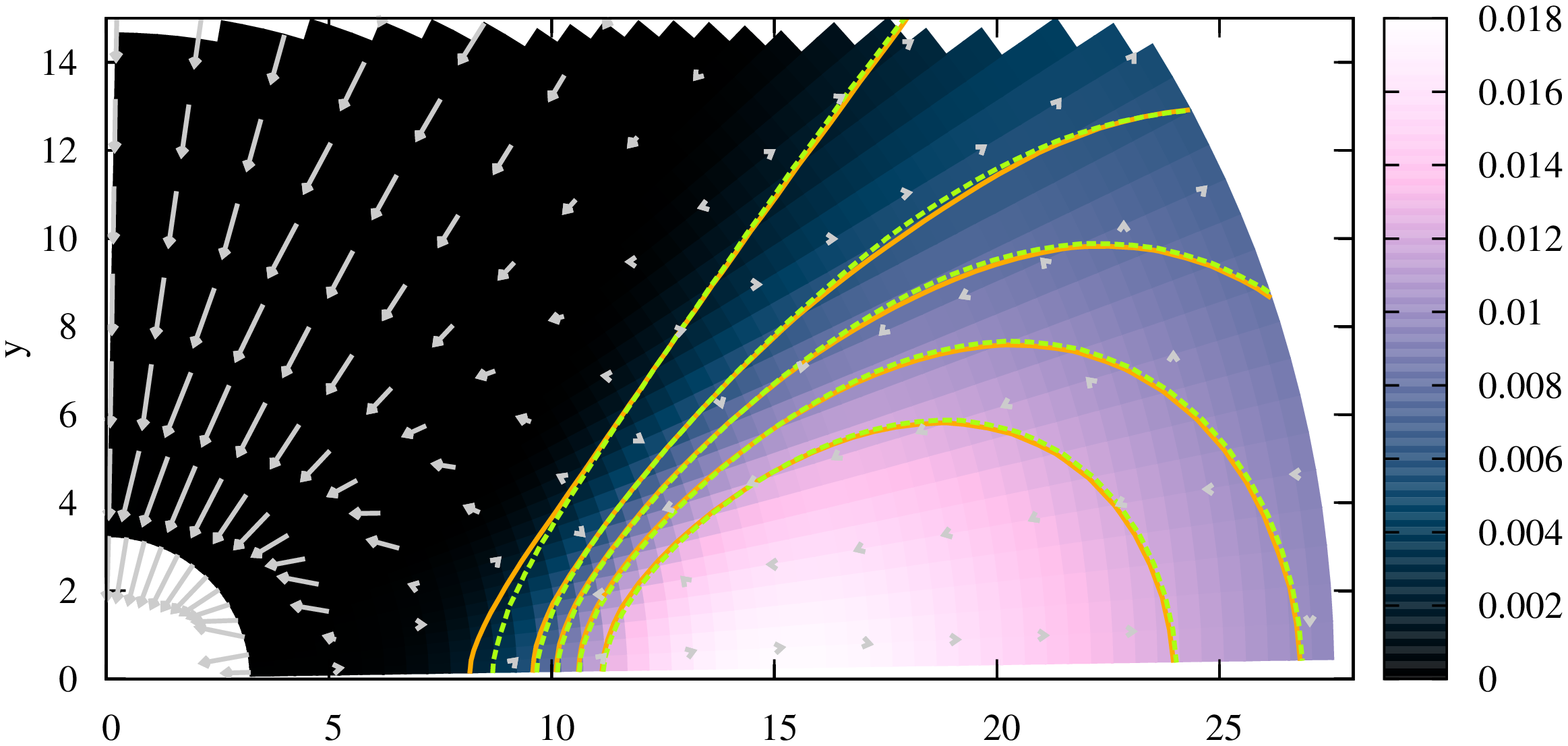}}\\\vspace{-.32in}
\subfigure{\includegraphics[width=.545\columnwidth,angle=270]{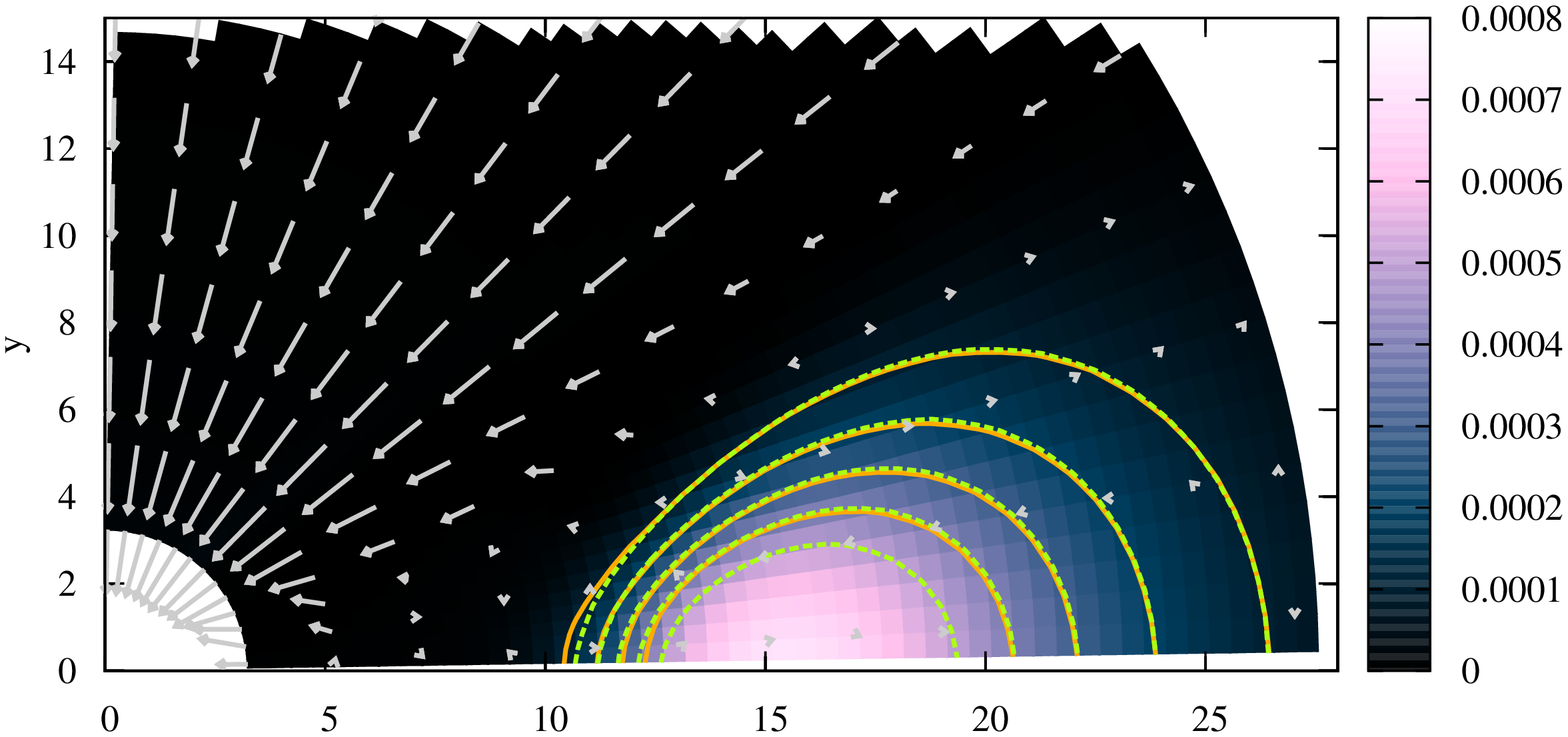}}\\\vspace{-.32in}
\subfigure{\includegraphics[width=.545\columnwidth,angle=270]{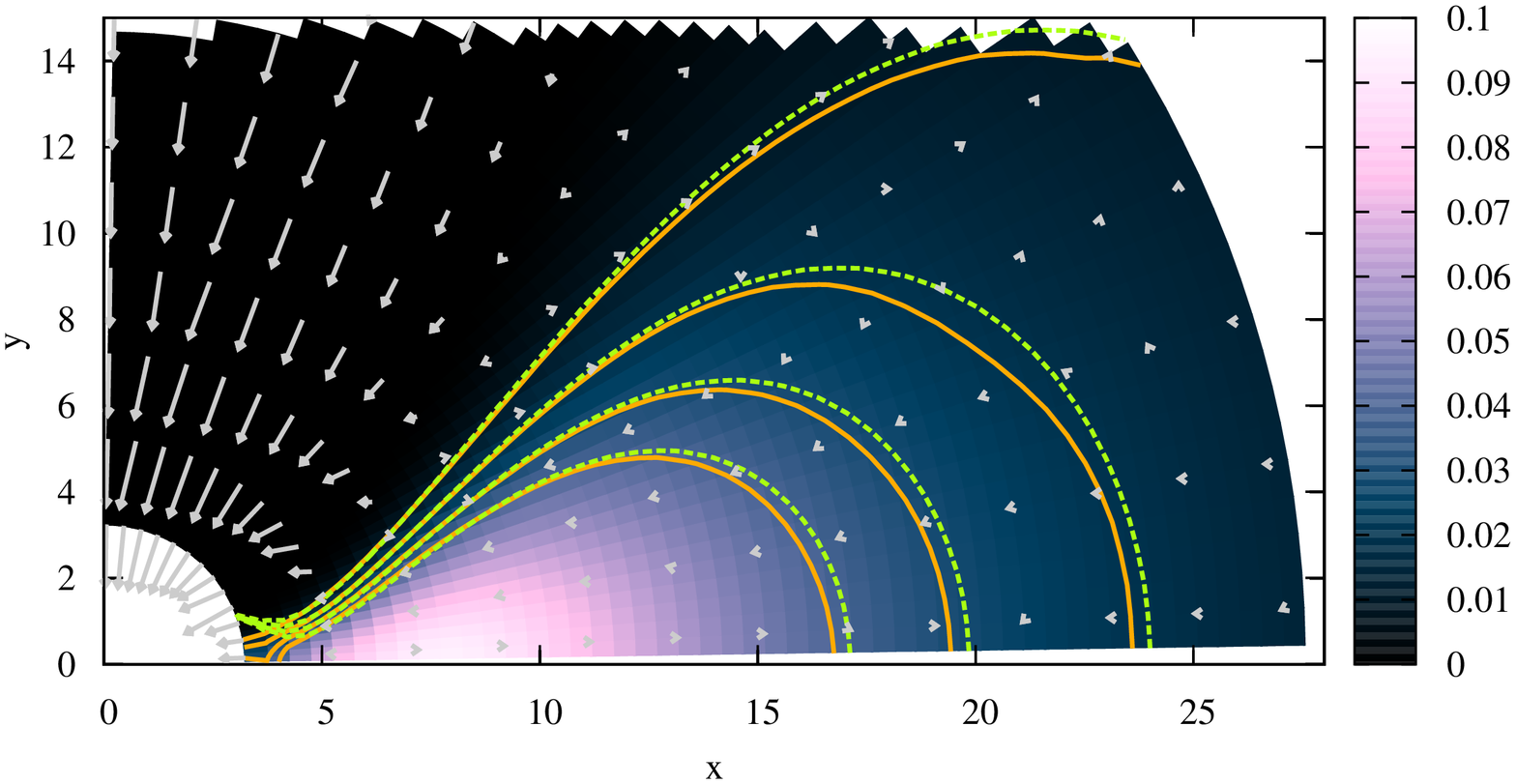}}
\caption{Equilibrium torii (Polish doughnuts) around a Schwarzschild
  BH after numerical evolution up to time $t=1000 M$. Colors and
  arrows show density and poloidal velocity, respectively. The green
  dashed lines show iso-density contours of the analytical
  solutions. Contours for the numerical solutions are shown with
  orange solid lines. The parameters of the three model torii are
  given in Table~\ref{t.doughnuts}.}
  \label{f.donut}
\end{figure}

Figure~\ref{f.donut} compares the relaxed, stationary solutions
obtained by running the code with the corresponding analytical Polish
doughnut solution. Colors and orange solid lines show the numerical
solution (density), while dashed green lines show the analytical
model. Vectors indicate the poloidal component of velocity.

The top panel in Figure~\ref{f.donut} corresponds to Model 1, where
the analytical solution extends from $r=8$ to the outer boundary. The
analytical and numerical contours agree very well. The only visible
discrepancy is in the low-density region near the torus inner edge,
where there is some numerical dissipation. 

The middle panel in Figure~\ref{f.donut} corresponds to Model 2, where
the analytical torus is entirely confined within the grid: $r_{\rm
  in}=10$, $r_{\rm out}=27$. Again, the numerical solution agrees very
well with the analytical solution except for the innermost region of
the torus where the numerical solution is slightly stretched inward.

The bottom panel in Figure~\ref{f.donut} shows Model 3. The analytical
contours plotted correspond to open equipressure surfaces which
intersect the BH horizon.  Non-zero poloidal velocities are thus
expected in this model and the analytical solution is not fully
consistent. Nevertheless, there is a close similarity between the
numerical and analytical solutions, which means that the code is able
to represent even this torus quite well.

\subsection{Radiative shock tubes}
The previous two tests did not involve radiation. For our first test
with radiation, we set up a number of radiative shock tube problems as
described in \cite{farrisetal08} and \cite{roedigetal12}. These shock
tubes are similar to the pure hydrodynamic shock tube described in
Section~\ref{s.hdtube}, i.e., the system begins with gas in two
different states (left and right), separated by a membrane. The
membrane is removed at $t=0$ and the system is allowed to evolve. The
difference here is that the evolution is described by the full set of
equations~(\ref{eq.hdlab1})--(\ref{eq.radlab1}). In addition, the
left- and right-states of all the tests except test No.~5 are set up
in such a way that the shock asymptotically becomes stationary
\citep[see Appendix C of][]{farrisetal08}.

Table~\ref{t.radtube} lists the parameters describing the initial
states of seven test problems that we have simulated. The scattering
opacity in all the tests is set to zero,
so $\chi=\kappa$ (equation~\ref{eq:kappaes}). 
The value of the radiative constant $\sigma$ in
code units is given in the table. All the tests were solved on a grid
of $800$ uniformly spaced points.  For consistency with previous work
by other authors, the Eddington approximation was used
(equation~\ref{eq.eddapr}).
\begin{table*}
\begin{minipage}{2\columnwidth}
\caption{Radiative shock tubes}
\label{t.radtube}
\begin{tabular}{@{}lcccccccccccccc}
\hline
Test & & & & & \multicolumn{4}{c}{Left state:}&  &\multicolumn{4}{c}{Right state:} \\
No. & $\Gamma$ &  $\sigma$ & $\kappa/\rho$ &  &$\rho$ & $p$ & $u^x$ & $\widehat E$ &  &  $\rho$ & $p$ & $u^x$ & $\widehat E$  \\
\hline
1 & 5/3 & $3.085\cdot10^9$ &0.4 & & 1.0 & $3.0\times 10^{-5}$ & 0.015 & $1.0\times10^{-8}$ & & 2.4 & $1.61\times10^{-4}$&$6.25\times10^{-3}$&$2.51\times10^{-7}$\\
2 & 5/3 & $1.953\cdot10^4$ &0.2 & & 1.0 & $4.0\times 10^{-3}$ & 0.25  & $2.0\times10^{-5}$ & & 3.11& $4.512\times10^{-2}$&$8.04\times10^{-2}$&$3.46\times10^{-3}$\\
3a& 2   & $3.858\cdot10^{-8}$&0.3&&  1.0 & $6.0\times 10^{1}$ & 10.0  & $2.0   $            & & 8.0& $2.34\times10^{3}$&$1.25$&$1.14\times10^{3}$\\
3b& 2   & $3.858\cdot10^{-8}$&25.0&&  1.0 & $6.0\times 10^{1}$ & 10.0  & $2.0   $            & & 8.0& $2.34\times10^{3}$&$1.25$&$1.14\times10^{3}$\\
4a& 5/3 & $3.470\cdot10^{7}$&0.08&&  1.0 & $6.0\times 10^{-3}$ & 0.69  & $0.18   $            & & 3.65& $3.59\times10^{-2}$&$0.189$&$1.3$\\
4b& 5/3 & $3.470\cdot10^{7}$&0.7&&  1.0 & $6.0\times 10^{-3}$ & 0.69  & $0.18   $            & & 3.65& $3.59\times10^{-2}$&$0.189$&$1.3$\\
5& 2   & $3.858\cdot10^{-8}$&1000.0&&  1.0 & $6.0\times 10^{1}$ & 1.25  & $2.0   $            & & 1.0& $6.0\times 10^{1}$&$1.10$&$2.0$\\
\hline
\end{tabular}
\end{minipage}
\end{table*}

\begin{figure}
  \centering
\includegraphics[width=.9\columnwidth,angle=270]{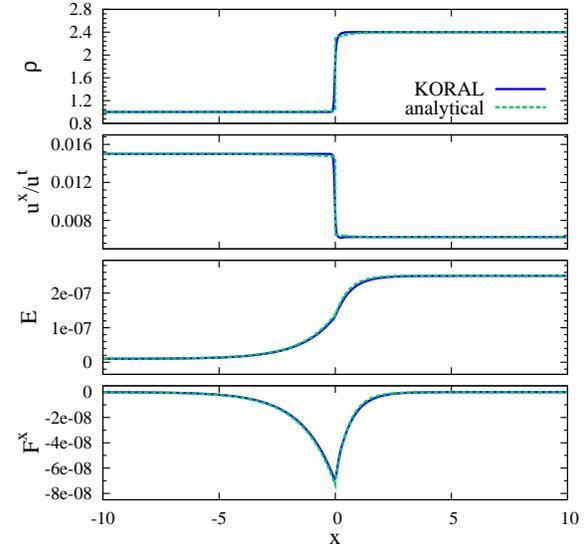}
\caption{Results obtained for radiative shock tube test No.~1. From
  top to bottom, the panels show the profiles of gas density, proper
  velocity, fluid-frame radiative energy density and fluid-frame radiative flux. The numerical
  solution obtained with \koral\ is indicated by solid lines and the
  semi-analytical solution given in \citet{farrisetal08} by dotted
  lines.
}
  \label{f.radtube1}
\end{figure}

Figure~\ref{f.radtube1} shows the numerical (solid) and analytical
(dashed lines) solutions for radiative shock tube problem No.~1, which
corresponds to a non-relativistic strong shock. The panels show (top
to bottom) density, proper velocity, fluid-frame radiative energy density and
fluid-frame flux, and may be directly compared (but for the flux which the other authors plot in the lab frame) to the corresponding figures and
analytical solutions of 
\citep{farrisetal08,zanottietal11, fragileetal12}. The agreement is
good.

\begin{figure}
  \centering
\includegraphics[width=.9\columnwidth,angle=270]{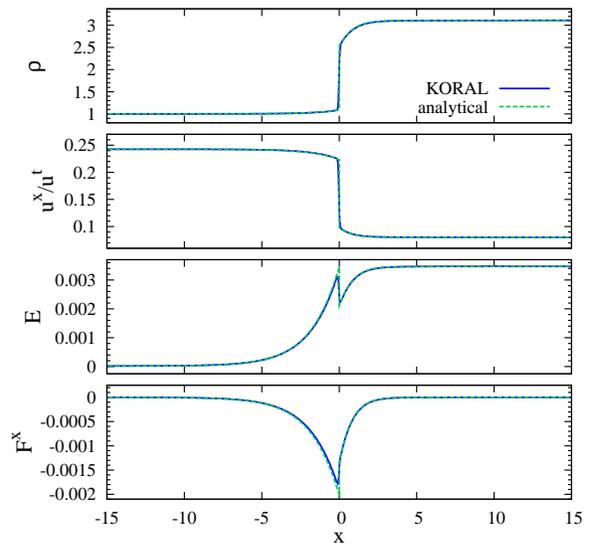}
\caption{Same as Figure~\ref{f.radtube1} but for radiative shock tube
  test No.~2.}
  \label{f.radtube2}
\end{figure}

Figure~\ref{f.radtube2} shows results for radiative shock tube test
No.~2, which corresponds to a mildly relativistic strong shock. Again,
the agreement between the numerical and semi-analytical
\citep{farrisetal08} profiles is good, except for a slight smoothing
of the numerical profiles at the position of the shock (see the bottom
panel). 

\begin{figure}
  \centering
\includegraphics[width=.9\columnwidth,angle=270]{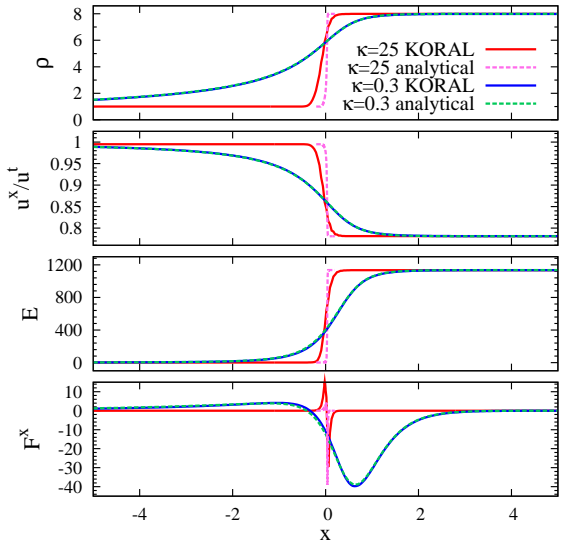}
\caption{Similar to Figure~\ref{f.radtube1} but showing results for
  radiative shock tube tests No. 3a (solid blue: numerical solution,
  dotted green: exact semi-analytical solution) and No. 3b (solid red:
  numerical, dotted magenta: semi-analytical).}
  \label{f.radtube3}
\end{figure}

Figure~\ref{f.radtube3} shows results corresponding to radiative shock
tube tests No.~3a and 3b. These are strongly relativistic shocks with
upstream $u^x=10$. Test No.~3a corresponds to shock tube test 3 of
\cite{farrisetal08}, while test 3b is the optically thick version of
the same test which was proposed and solved by
\cite{roedigetal12}. These two tests verify that the code is able to
resolve a highly relativistic wave in two very different optical depth
limits. In both cases, the numerical solution reaches a steady state
and closely follows the corresponding semi-analytical solution. The
only noticeable difference is that in the high-opacity case the
discontinuity in the numerical solution is less steep than in the
exact analytical profile. This discrepancy measures the amount of
numerical dissipation introduced by the algorithm.

\begin{figure}
  \centering
\includegraphics[width=.9\columnwidth,angle=270]{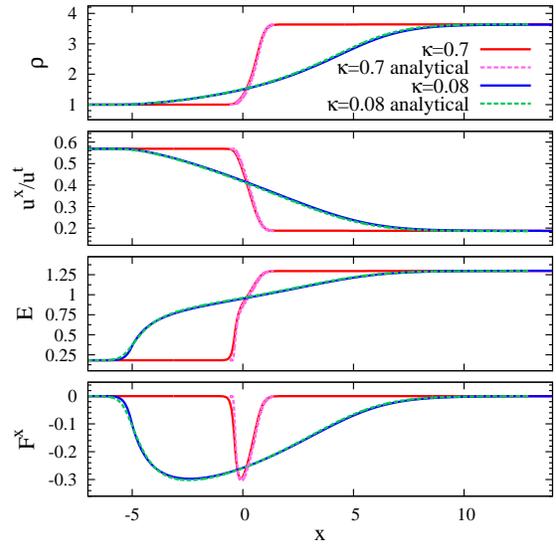}
\caption{Similar to Figure~\ref{f.radtube1} but showing results for
  radiative shock tube tests No. 4a (solid blue: numerical solution,
  dotted green: semi-analytical solution) and No. 4b (solid red:
  numerical, dotted magenta: semi-analytical).}
  \label{f.radtube4}
\end{figure}

Figure~\ref{f.radtube4} shows results for radiative shock tube tests
No.~4a and 4b. These tests correspond to radiation pressure dominated
mildly relativistic waves. Test~4b is the optically thick version of
test~4a that was proposed by \cite{roedigetal12}. In both tests, the
numerical solution reaches a stationary state and agrees well with the
semi-analytical solution.  Note that the values of the opacity
coefficient $\kappa$ in tests 3b and 4b are the maximum values that
the numerical scheme of \cite{roedigetal12} could handle. The
algorithm implemented in \koral\ has no such a limitation. We could
increase $\kappa$ to much larger values and the scheme would remain
stable.

\begin{figure}
  \centering
\includegraphics[width=.9\columnwidth,angle=270]{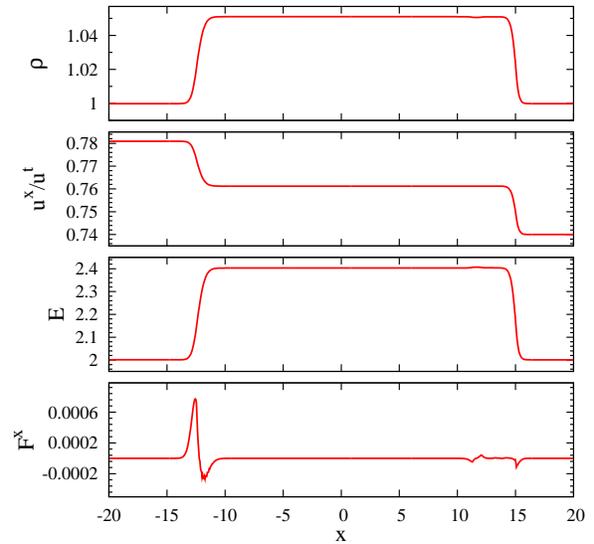}
\caption{Same as Figure~\ref{f.radtube1} but for radiative shock tube
  test No.~5. There is no analytical solution available for this problem.
}
  \label{f.radtube5}
\end{figure}

Figure~\ref{f.radtube5} corresponds to radiative shock tube test
No.~5.  This is the only test that does not asymptote to a stationary
solution. This test was proposed and solved by \citet{roedigetal12}
and represents an optically thick flow with mildly relativistic
velocities. The left- and right-states are identical except that they
have different velocities. As a result, two shock waves propagate in
opposite directions. This test does not have an analytical
solution. However, by comparing our numerical solution with that
presented in \cite{roedigetal12}, we confirm that our scheme
performs satisfactorily.

\subsection{Radiative pulse}
 \label{s.diffusion}

 We now test the ability of our scheme to handle the evolution of a radiation pulse in
 the optically thick and optically thin limits. We start with the optically thin case.
We set up a Gaussian distribution of radiative energy density at the center of a 3D Cartesian coordinate system. The pulse radiative temperature is
 set according to, \be
 T_{\rm rad}=\left(\frac{E}{4\sigma}\right)^{1/4}=T_0\left(1+100 e^{-(x^2+y^2+z^2)/w^2}\right),
 \ee with $T_0=10^6$, $w=5.0$ and the radiative constant $\sigma=1.56\times10^{-64}$. We assume zero absorption opacity ($\kappa=0$) and non-zero but negligible scattering opacity ($\kappa_{\rm es}=10^{-6}$). The background fluid field has constant density $\rho=1$ and temperature $T=T_0$. We solve the problem in three dimensions on a coarse Cartesian grid of $51x51x51$ cells using the linear reconstruction with $\theta=2$.

\begin{figure*}
   \centering
   \subfigure{\includegraphics[height=.343\textwidth,angle=270]{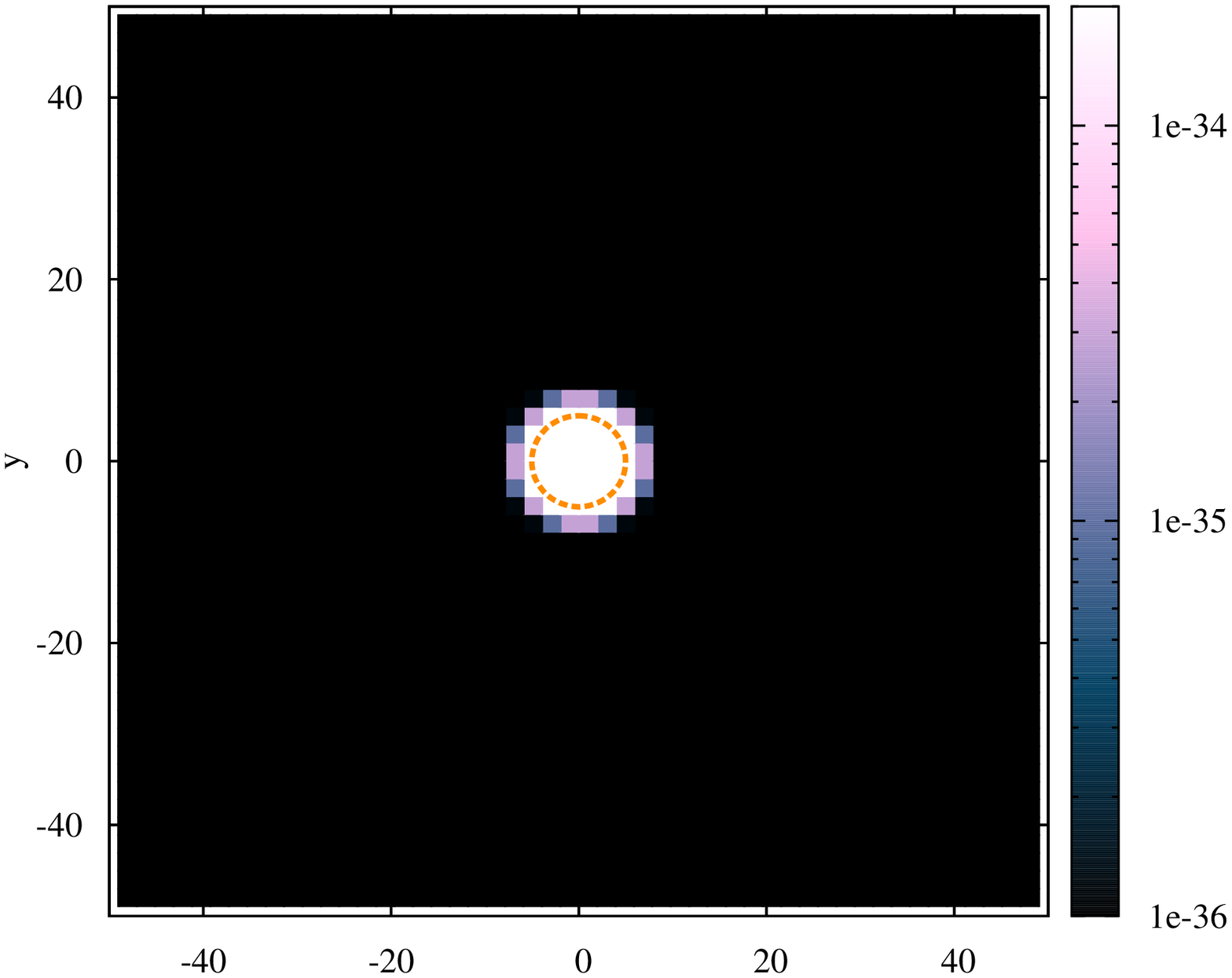}}\hspace{-.75cm}
   \subfigure{\includegraphics[height=.343\textwidth,angle=270]{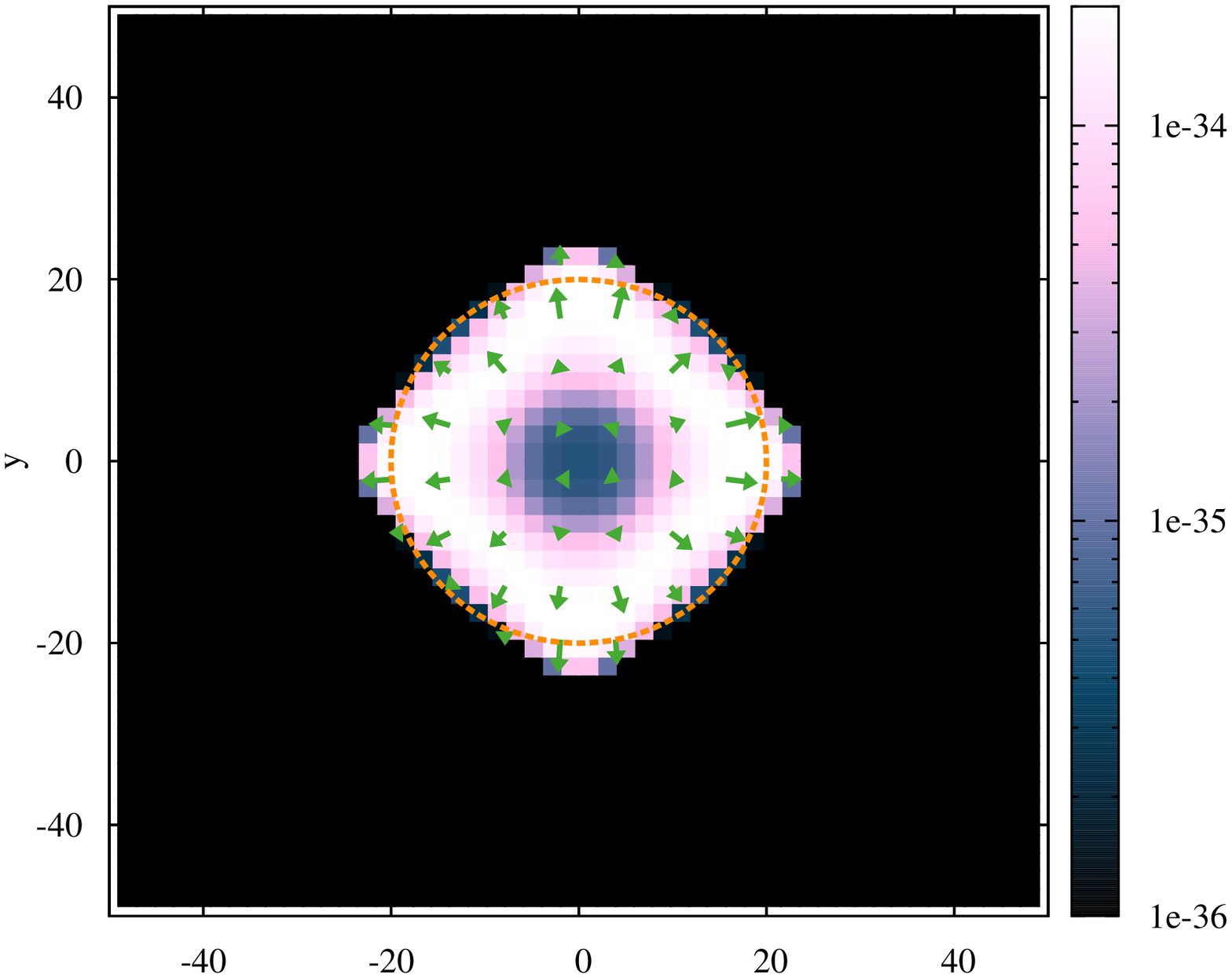}}\hspace{-.75cm}
   \subfigure{\includegraphics[height=.343\textwidth,angle=270]{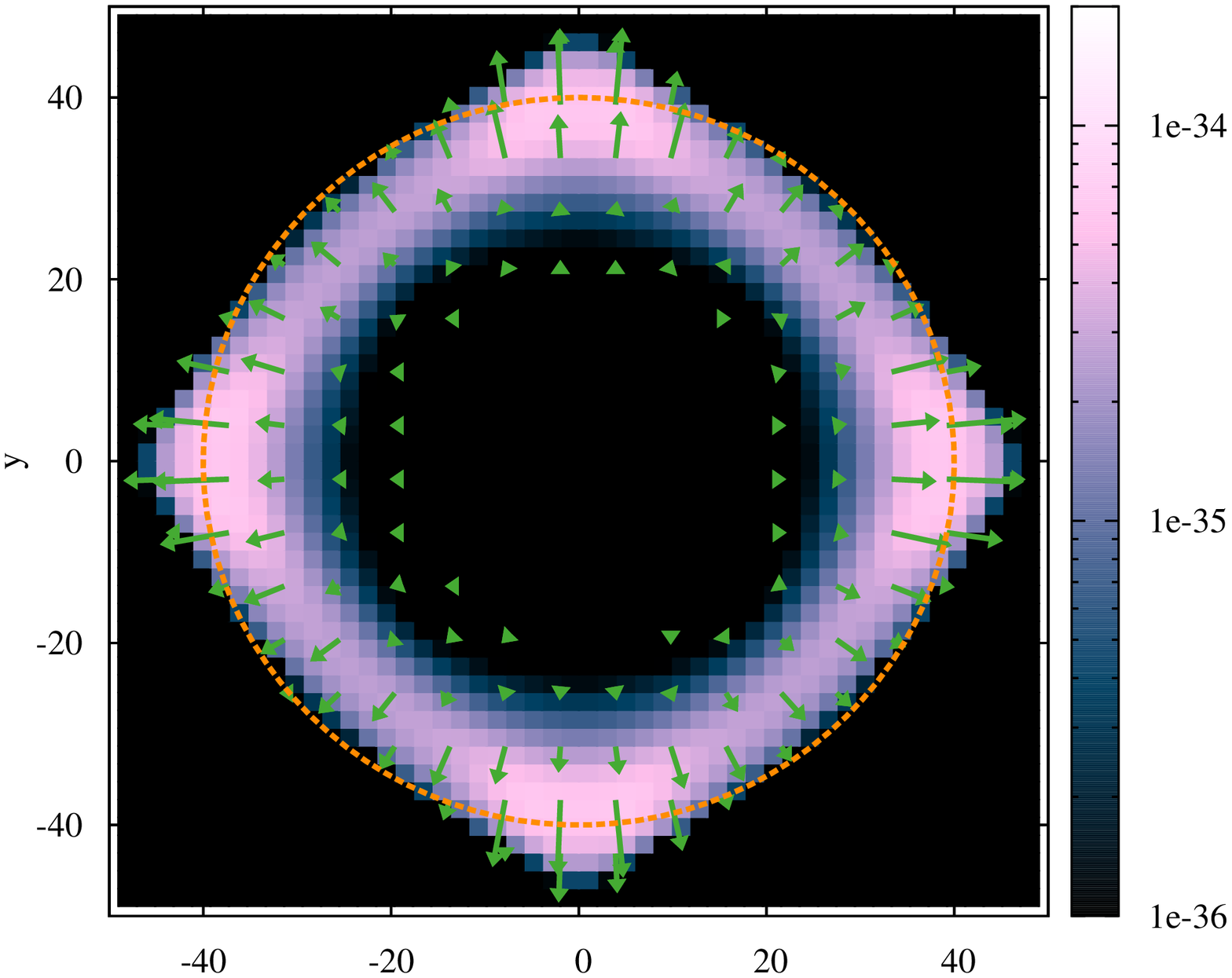}}\vspace{-.75cm}\\
   \subfigure{\includegraphics[height=.2813\textwidth,angle=270]{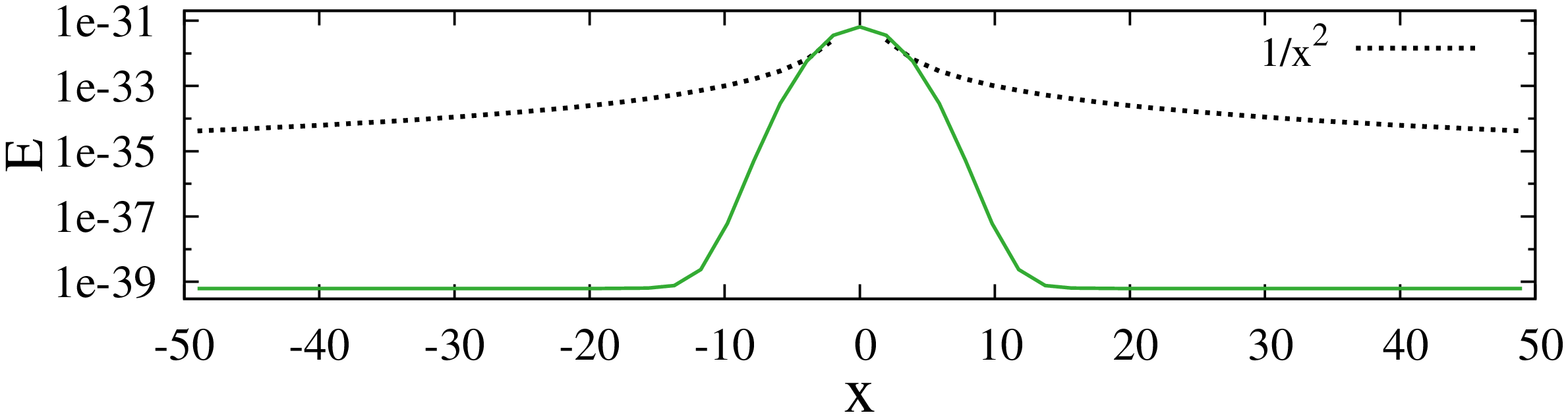}}\hspace{.3385cm}
   \subfigure{\includegraphics[height=.2813\textwidth,angle=270]{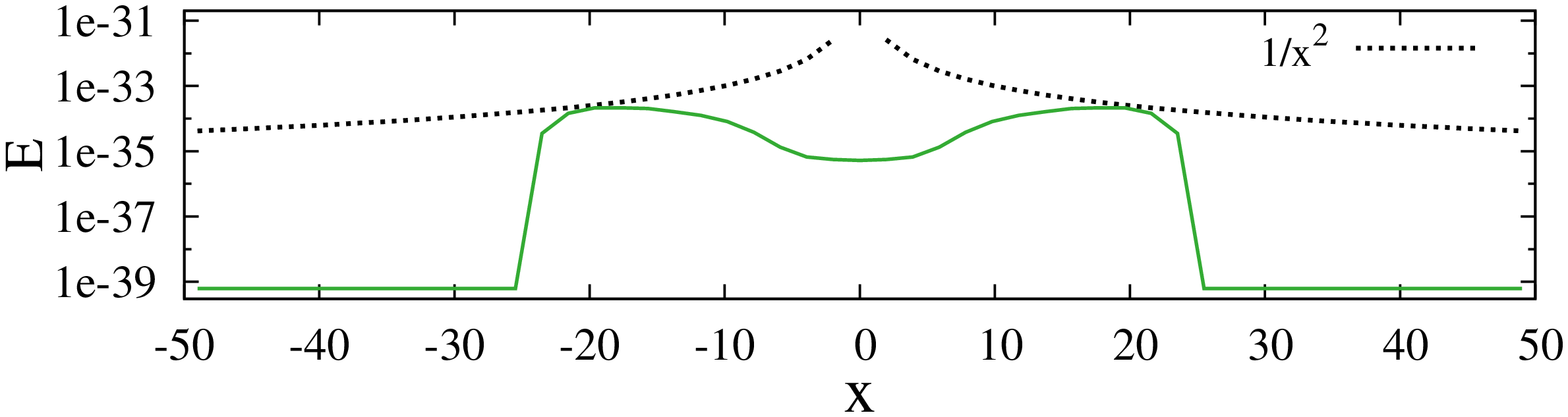}}\hspace{.3385cm}
   \subfigure{\includegraphics[height=.2813\textwidth,angle=270]{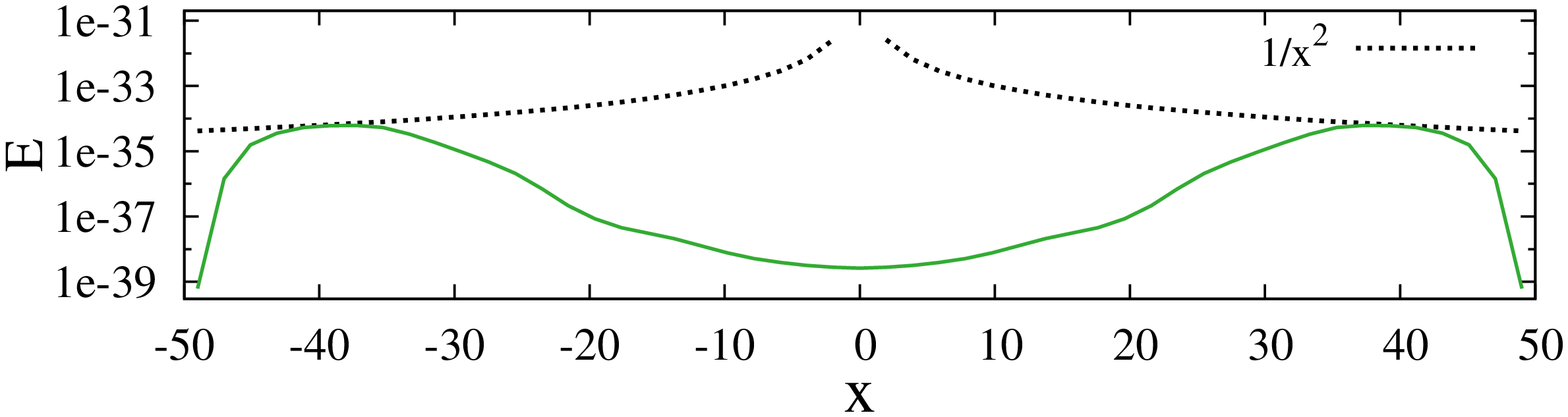}}\hspace{.87cm}
   \caption{    
     Profiles of the radiative energy density ($E$) for the optically thin radiative pulse test described in
     Section~\ref{s.diffusion}. The top panels show its distribution in the $xy$ plane at (from left to right) $t=0$, $15$ and $35$. The orange circle in the first plot denotes the initial width and expands at the speed of light to provide the expected pulse front location in the other two plots. The bottom panels show the corresponding profiles measured along $y=z=0$ line and the $1/x^2$ dependence}
   \label{f.pulsethin}
 \end{figure*}

The initial pulse in radiative energy density is expected
 to spread isotropically with the speed of light (optically thin medium) and to decrease inversely proportionally to the square of radius (energy conservation). Such behavior is visible in Fig.~\ref{f.pulsethin} showing the radiative energy distribution in the $z=0$ plane (top panels) and its cross-section along $y=z=0$ (bottom panels). The orange circles in the top set of panels show the expected size of the pulse. It is clear that the propagation speed of the pulse is consistent. This problem was solved on a relatively coarse Cartesian grid. This results in deviations from the perfectly spherical shape --- the radiative energy density of the pulse along the axes is higher than along the diagonals. This effect is reduced by choosing larger resolution or a more suitable grid (e.g., spherical). The bottom set of panels in Fig.~\ref{f.pulsethin} shows the profiles of the energy density along the $x$-axis. The black dotted lines show the expected rate of energy decrease with increasing distance from the center. The numerical solution perfectly follows this trend.

 To test the optically thick limit we choose to set up a similar pulse but this time planar instead of a point-like, i.e., according to,
\be
 T_{\rm rad}=\left(\frac{E}{4\sigma}\right)^{1/4}=T_0\left(1+100 e^{-x^2/w^2}\right).
 \ee 
This time we set the scattering opacity to $\kappa_{\rm es}=10^{3}$ and solve the problem as one-dimensional on $101$ grid points distributed uniformly between $x=-50$ and $x=50$ with periodic boundary conditions in $y$ and $z$. The total optical depths per cell and per pulse are therefore $\tau=10^3$ and $\tau=10^4$, respectively.

In the optically thick limit the evolution of such a system is described by a diffusion equation,
\be
\partial_t E=\frac{1}{3\chi}\partial_{xx}E,
\label{eq.difflimit}
\ee
which can be derived from the non-relativistic limit of Eq.~\ref{eq.radlab1} assuming the time derivative of the $x$ component of the flux vanishes. An initially Gaussian pulse of radiative internal energy will therefore diffuse according to the value of the diffusion coefficient $1/3\chi$.

\begin{figure}
  \centering
\includegraphics[width=.95\columnwidth,angle=270]{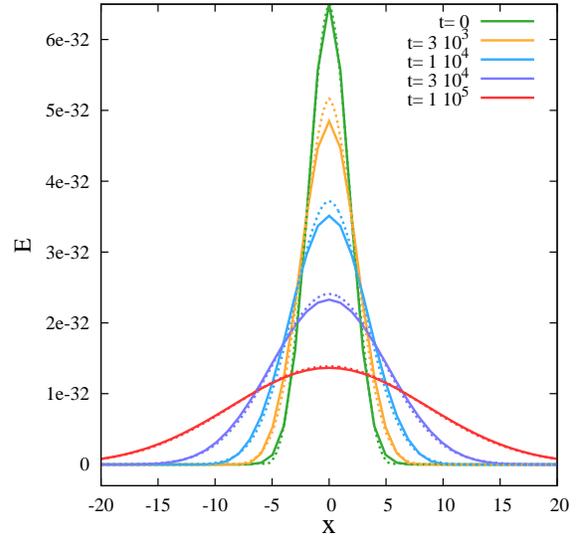}
\caption{The radiative energy density for the optically thick pulse described in Section~\ref{s.diffusion}. The green line shows the initial profile while the other color show the time evolution. Dotted lines show the exact solution of Eq.~\ref{eq.difflimit}.}
  \label{f.pulsethick}
\end{figure}

In Fig.~\ref{f.pulsethick} we plot profiles of the radiative energy at various moments (solid lines) and compare them to the exact solution of Eq.~\ref{eq.difflimit} (dotted lines). The numerical solution diffuses slightly faster due to the additional numerical dissipation introduced by the scheme\footnote{This artificial numerical diffusion may be further reduced by stronger damping of the characteristic radiative wavespeed in the optically thick limit (Eq.~\ref{eq.wavespeedlimit}). However, it might cause problems in the intermediate regime.}. At later times this difference becomes insignificant.

\subsection{Shadow test}
\label{s.shadow}

Here we test the ability of the M1 closure scheme, as incorporated in
\koral, to resolve shadows. We set up a blob of dense, optically
thick gas in flat spacetime, surrounded by an optically thin medium,
and we illuminate this system.

We start with a single source of light imposed on the left boundary. We
solve the problem in two dimensions on a 100x50 grid, with the density
distribution set to be
\be
\rho = \rho_0+(\rho_1-\rho_0)\,e^{-\sqrt{x^2+y^2}/w^2},
\ee
where $\rho_0=10^{-4}$, $\rho_1=10^3$ and $w=0.22$. The gas
temperature is adjusted so as to give constant pressure throughout
the domain,
\be
T = T_0\frac\rho{\rho_0}.
\ee

The initial radiative energy density is set to the local thermal
equilibrium value, and the initial velocities and radiative fluxes are
zero. We apply periodic boundary conditions at the top and bottom and
outflow boundary conditions at the right border of the domain. At the
left border we have the external source of light, which we specify
with $E_L=4\sigma T_L^4$, $F^x=0.99999E_L$, $T_L=100T_0$. All other
quantities are set to match the ambient gas. We evolve the system with
both the Eddington approximation and M1 closure, assuming
$\kappa=\chi=\rho$.

\begin{figure}
  \centering
\includegraphics[width=.8\columnwidth,angle=270]{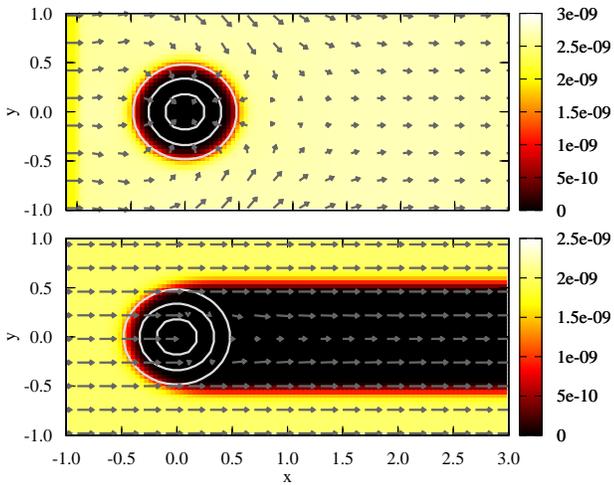}
\caption{Results obtained with the shadow test for a single beam of light. The upper panel
  corresponds to the Eddington approximation and the lower panel to
  the M1 closure scheme. Colors denote radiative energy density,
  contours show density ($\rho=1$, $100$, $500$) and arrows show the
  direction and magnitude of the reduced radiative flux $F^i/E$. }
  \label{f.shadow}
\end{figure}

Figure~\ref{f.shadow} shows the results at $t=10$. By this time, the
initial radiation wave has passed through the domain and the system
has reached a stationary state. The upper panel shows the solution we
obtain with the Eddington approximation. Because this closure treats
all directions equally, radiation readily diffuses into the region
behind the blob. As a
result, there is no shadow behind the optically thick blob. The lower
panel in Figure~\ref{f.shadow} shows results with the the
M1-closure. In contrast to the case of Eddington closure, here the $x$
direction is distinguished because the incoming radiation at the left
boundary moves in this direction. The M1 closure is designed to keep
flux moving parallel to itself in optically thin regions for $F\approx E$. As a result,
a strong shadow develops behind the optically thick blob. This test
illustrates one of the key differences between the Eddington and M1
closure schemes.

It is appropriate to mention that the excellent performance of the M1
closure scheme in this shadow test problem is partly because the setup
is particularly favorable. First, we have only a single source of
radiation. Second, the shadow here is
aligned well with the grid, which helps to minimize diffusion.  For
other grid orientations, the numerical results would exhibit more
diffusion, e.g., see Section~\ref{s.beamBH}.

\begin{figure*}
  \centering
\includegraphics[height=.8\textwidth,angle=270]{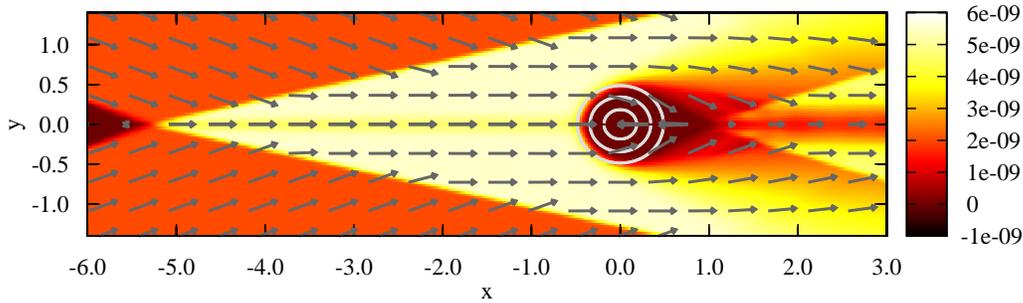}
\caption{Similar to Fig.~\ref{f.shadow} but for two inclined beams of light. }
  \label{f.shadow2}
\end{figure*}

The M1 closure assumes that the specific intensity is symmetric with respect to the radiative flux, i.e., only one direction is distinguished. It means that this approach is supposed to be less efficient when multiple sources of light are involved. To test its performance for such a setup we implemented a two-beam test problem similar to the one described in
\citet{jiangetal12}\footnote{The key principle behind the non-relativistic algorithm described in
\cite{jiangetal12} is the use of a  ``Variable Eddington Tensor''
(VET). The VET is used to close the radiative equations, relating
radiative pressure to the local radiative energy density. The VET is
computed through a separate radiative transfer solver, which
calculates (at each time step) the time-independent radiation field
(using the fixed fluid background of the previous timestep as its
input).  The authors solve the radiative transfer equations along a
discrete set of rays, and so their scheme accurately captures all
shadows that can be resolved by these ray angles.  The radiation
pressure obtained from the VET is then used to evolve the MHD fluid
equations.}. We set up exactly the same initial conditions for gas and radiation as in the two previous tests. This time, however, we set up a reflection symmetry at the lower boundary ($y=0$) and we impose an inclined ($F^x_0=0.93E_0$, $F^y_0=-0.37E_0$) beam on the upper boundary and on the part ($y>.3$) of the left boundary. As a result, the domain is effectivelly enlighted by two self-crossing beams of light. We plot the result of a numerical simulation in Fig.~\ref{f.shadow2} where the region of negative $y$-coordinates was plotted by reflecting the $y$-positive data. In the region near the left top and bottom corners, where the beams do not overlap, the direction of the flux follows the imposed boundary condition. In the region of the overlap the radiative energy density increases twice ($E=2E_0$) while the flux becomes equivalent to the superposition of the beam-intrinsic fluxes, i.e., it is purely horizontal and its $x$-component equals $F^x=2F^x_0=1.86E_0=0.93E$. The clump of optically thick gas is, therefore, effectively illuminated by a purely horizontal beam. One could expect it creates a parallel shadow similar to the one obtained in the single beam problem. This is, however, not the case. There is an important difference between the beam we imposed on the left boundary in the single beam problem and the one which develops in the overlapping region. The former had $F^x\approx E$ what implied that the specific intensity was almost a $\delta$ function in the direction of the flux. The latter, however, has $F^x=0.93E$ which means that the implied distribution of the specific intensity is only an elongated ellipsoid pointing in the direction of the flux, i.e., photons moving in other directions than the direction of the flux are allowed. This has an effect on the shadow produced behind the clump. Instead of sharp parallel edges we get regions of the partial shadow (penumbra) resulting from these perpendicular photons allowed by the closure when $F^x<E$. The region of the total shadow (umbra) is therefore limited by the edges of the penumbra and follows the expected shape (compare Fig.~11 in \citet{jiangetal12}) to a good accuracy. A significant difference between the exact solution and our numerical one arises in the region where the penumbrae overlap. One could expect a uniform, triangular region of no shadow. The M1 closure, however, produces an extra narrow horizontal shadow along the $x$-axis. 

This test shows limits of the M1 closure approach but at the same time stresses the fact that, in principle, it does not limit specific intensity to one particular direction (assuming only its symmetry with respect to the flux). It performs much better than the Eddington approximation but in the case of multiple sources of light it must be used with caution.

\subsection{Static atmosphere}
\label{s.atmosphere}

An important aspect of radiation in accretion disks is momentum
transfer between radiation and gas. The Eddington luminosity limit,
for instance, arises from this interaction.  To validate the treatment
of gas-radiation momentum exchange in our method, we study a static
atmosphere which is in equilibrium under the combined action of
gravity, gas pressure gradient and radiation force. We consider a
polytropic atmosphere on a spherical object. We take the optically
thin limit and assume that gas-radiation interactions occur only
through a scattering coefficient, i.e., $\kappa=0$, $\chi=\kappa_{\rm
  es}$ (equation~\ref{eq:kappaes}).

An analytical solution is easily derived for this model problem.
Because we assume a polytropic equation of state and set $\kappa=0$,
there is no energy equation, and the radial component of the
momentum equation ($r$ component of equation~\ref{eq.hdlab2}) is all
that matters. In the non-relativistic limit ($r\gg2$), assuming
stationarity ($\partial_t=0$) and zero velocity ($v^i=0$), this
equation takes the form 
\be\label{eq.atm1}
\frac1\rho\pder{p}{r}=-\frac{1-f}{r^2}, 
\ee
where
\be 
f=\kappa_{\rm es}F_{\rm in}r_{\rm in}^2.
\ee 
Here $F_{\rm in}$ is the radiative flux imposed as a boundary condition at
the bottom of the atmosphere, $r=r_{\rm in}$, and $f$ gives the ratio of
the radiative to gravitational (or geometrical) forces; $f=1$
corresponds to the Eddington limit, where the luminosity is
$L_{\rm Edd}=4\pi/\kappa_{\rm
  es}$ and the radiative flux is $F_{\rm in} = F_{\rm Edd} = 1/\kappa_{\rm
  es}r_{\rm in}^2$. Since radiative energy must be conserved, 
in the stationary state the flux must satisfy $F=F_{\rm in}r_{\rm in}^2/r^2$
(non-relativistic limit).

Equation~(\ref{eq.atm1}) may be solved with the help of the polytropic
equation of state $p=K\rho^\Gamma$ to give, 
\be\label{eq:rho}
\rho=\left[\frac{(\Gamma-1)}{\Gamma
  K}\left(C+\frac{1-f}r\right)\right]^{\frac1{\Gamma-1}}, 
\ee
where
\be\label{eq:f}
C=\frac{\Gamma K}{(\Gamma-1)}\rho_{\rm in}^{\Gamma-1} -
\frac{1-f}{r_{\rm in}}, 
\ee
and $\rho_{\rm in}$ is the assumed density
at $r=r_{\rm in}$. The entropy constant $K$ is calculated at the
bottom of the atmosphere from the assumed gas temperature $T_{\rm in}$.

We set up a linear grid of $40$ points between $r=10^6$ and $1.4\times10^6$
gravitational radii and we solved the problem using MP5 reconstruction
scheme and the standard M1 closure. We scaled all quantities to
physical units assuming $M=1M_\odot$ and $\kappa_{\rm es}=0.4\rho\,
{\rm cm^{-1}}$. At the innermost radius we set $\rho_{\rm
  in}=10^{-15}\, {\rm g\,cm^{-3}}$ (optically thin atmosphere) and $T_{\rm
  in}=10^6 \,\rm K$. All the velocities were initially zero and the
radiative energy density $E=F_{\rm in}/0.99$. Initial values of
the gas density and temperature in the domain and in the ghost cells were assigned based
on the analytical solution. We ran four models corresponding to four
luminosities: $10^{-10}$, $0.1$, $0.5$ and $1.0\, L_{\rm Edd}$.  Each
model was run up to a time $t=2\times10^9 M$, which is sufficient to
reach relaxed steady state for these optically thin atmospheres.

\begin{figure}
  \centering
\subfigure{\includegraphics[height=.473\textwidth,angle=270]{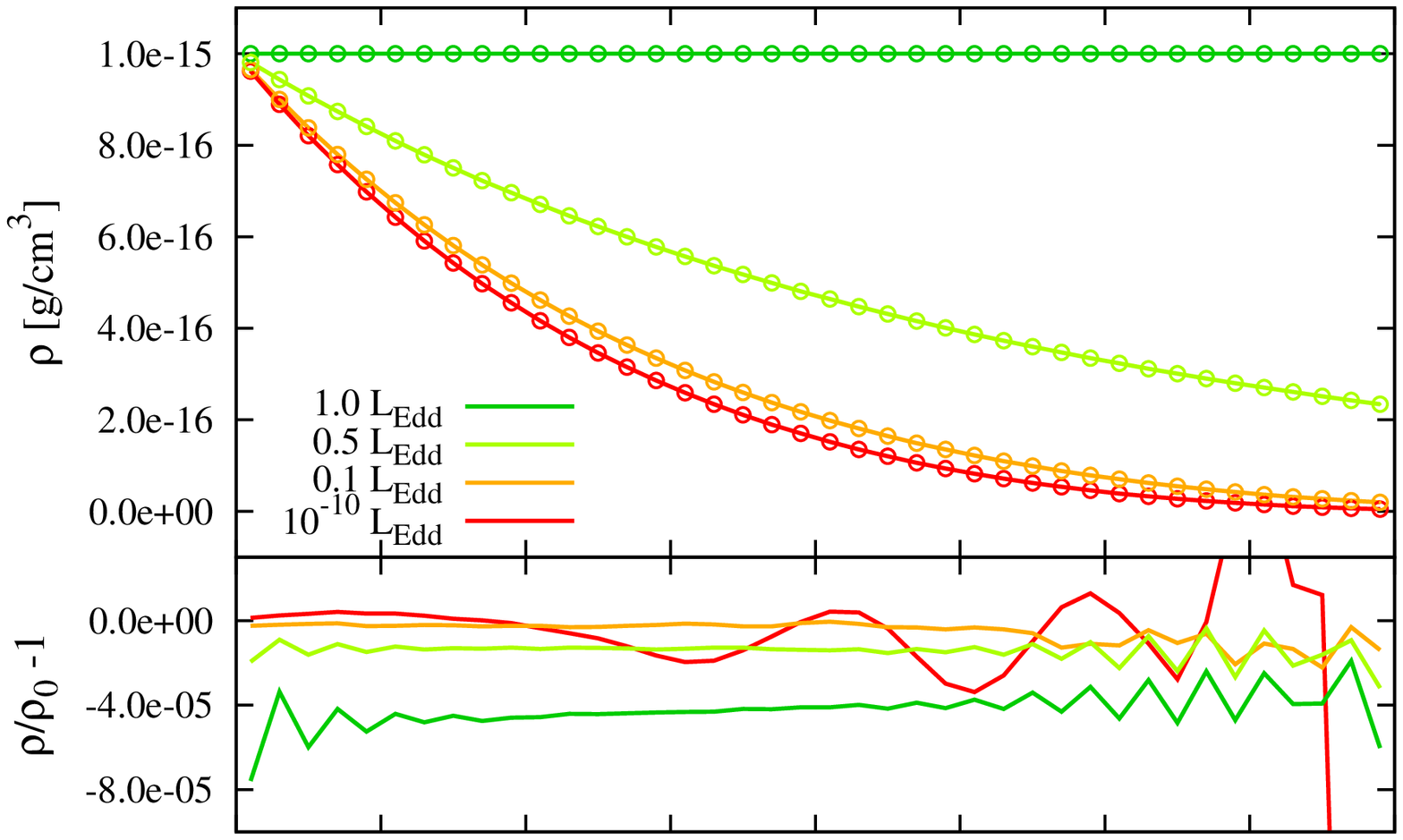}}\vspace{-.3in}
\subfigure{\includegraphics[height=.473\textwidth,angle=270]{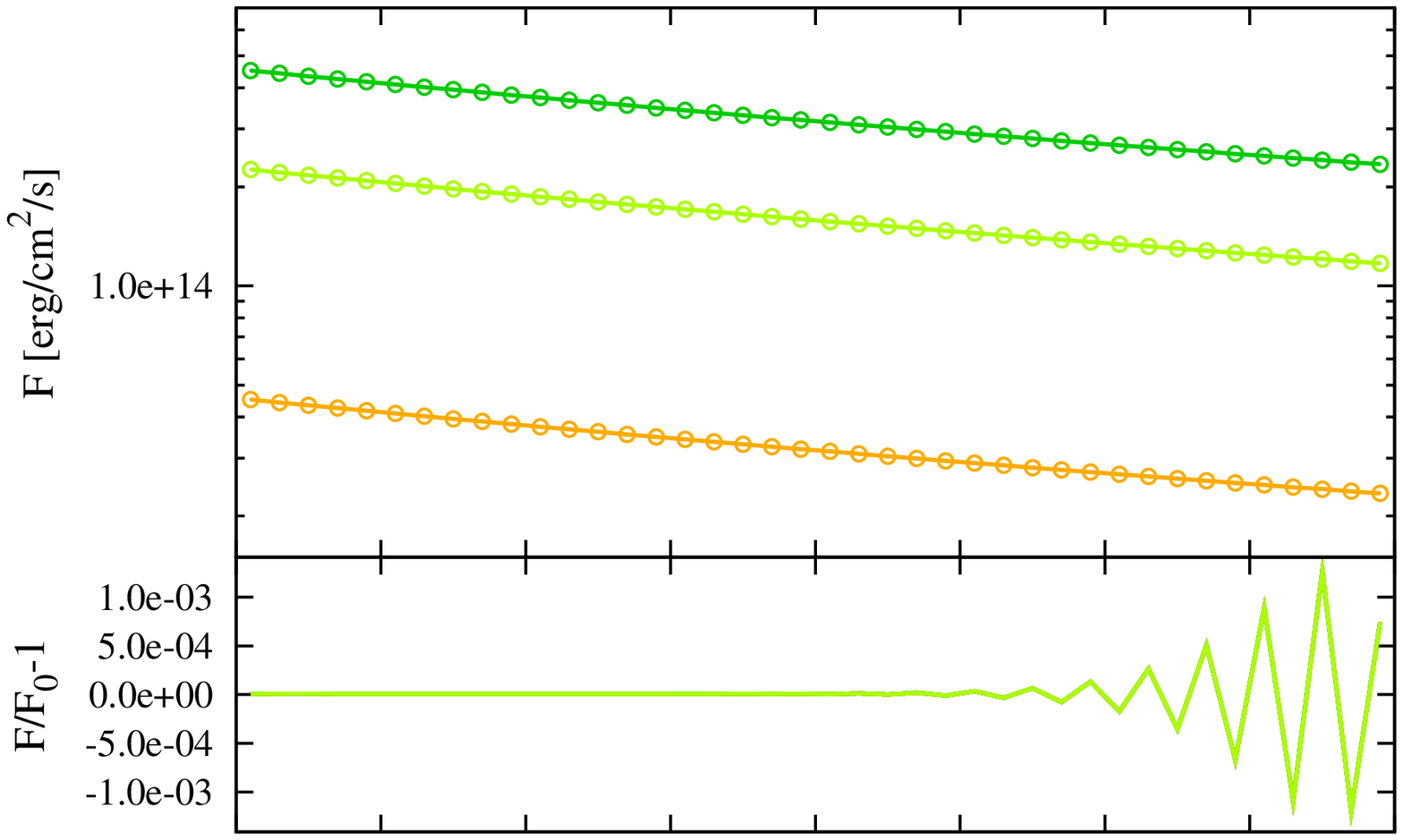}}\vspace{-.3in}
\subfigure{\includegraphics[height=.473\textwidth,angle=270]{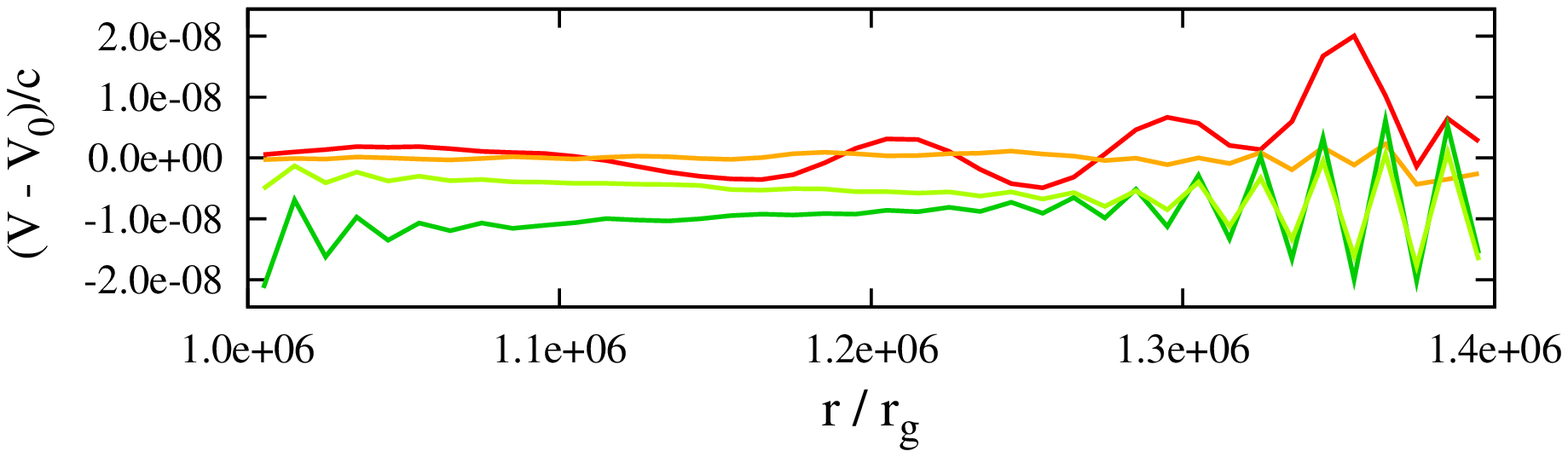}}
\caption{Results obtained with the static atmosphere test. Numerically
  determined profiles and residuals between the numerical and
  analytical solutions are plotted for the density (top panel), radial
  flux (middle panel) and radial velocity (bottom panel, residuals
  only). Colors denote the Eddington ratio of the flux boundary
  condition $F_{\rm in}$ at the bottom of the atmosphere: $F_{\rm in}
  = 10^{-10} F_{\rm Edd}$ (red), $0.1 F_{\rm Edd}$ (orange), $0.5
  F_{\rm Edd}$ (light green) and $1.0 F_{\rm Edd}$ (dark
  green) 
 Circles correspond to the numerical solutions and lines show the analytical profiles
 (equation~\ref{eq.atm1}).}
  \label{f.radatm}
\end{figure}

Figure~\ref{f.radatm} shows the results.  The top panel shows the
density profiles corresponding to the four models. Solid lines
are the analytical solutions and filled circles correspond
to the numerical solutions. The agreement is very good. The higher the
luminosity, the flatter is the density profile, indicating the effect
of the outward force due to radiation. For the particular case of the
Eddington luminosity, the density is perfectly constant, reflecting
the fact that the gravitational force is exactly balanced by radiation
and no pressure gradient in required. We see that the relaxed
numerical solution is indistinguishable from the analytical solution,
confirming that \koral\ properly handles gas-radiation momentum
exchange. The plot of residuals at the bottom of the panel indicates
that fractional deviations in the density are below $10^{-4}$. Even
this small discrepancy is at least in part because we are comparing a
numerical solution from a GR code with an analytical solution derived
under Newtonian physics.

The middle panel in Figure~\ref{f.radatm} shows our results for the
radial radiative flux.  Once again, the models behave very well and
the agreement with the analytical solution is excellent.  Finally, the
bottom panel shows the residual radial velocities ($v_r/c$). These are
of the order of $10^{-8}$ (they should be zero), and appear to be
mostly driven by slight inconsistencies near the boundaries (possibly
again because of a slight tension between GR and Newtonian physics).


\subsection{Beam of light near BH}
\label{s.beamBH}

To test the performance of the code for radiation in strong
gravitational field, we study propagation of a beam of light in the
Schwarzschild metric. We consider three models, in each of which a
beam of light is emitted in the azimuthal direction at a different
radius. We decouple gas and radiation by neglecting
absorptions and scatterings ($\kappa=\chi=0$). We run the models on a
two-dimensional grid with 30 points distributed uniformly in $r$
between $r_{\rm in}$ and $r_{\rm out}$ (see Table~\ref{t.beam} for
values) and 60 points distributed uniformly in azimuthal angle $\phi$ between
$\phi=0$ and $\pi/2$. 
Initially, we assign negligibly small values
for all primitive quantities, including the radiation energy density
and flux. We use outflow boundary conditions on all borders except the
region covered by the beam at the equatorial plane (see the range of
$r_{\rm beam}$ in Table~\ref{t.beam}), where we set the radiation
temperature to $T_{\rm beam}=10^{10}=1000T_0$ and the flux to
$F^\phi=0.99999E$. Here $T_0$ is the initial gas and radiation
temperature of the ambient medium. We use linear reconstruction
with $\theta=1$.

\begin{table}
\caption{Model parameters for the light beam tests}
\label{t.beam}
\centering\begin{tabular}{@{}lccc}
\hline
 Model & $r_{\rm beam}$ & $r_{\rm in}$ & $r_{\rm out}$\\
\hline
1 & $3.0\pm0.1$ & 2.5 & 3.5 \\
2 & $6.0\pm0.2$ & 5.3 & 7.5 \\
3 & $16.0\pm0.5$ & 14.0 & 20.5 \\
\hline
\end{tabular}
 \end{table}

\begin{figure*}
  \centering\hspace{-.15in}
\subfigure{\includegraphics[width=.35\textwidth,angle=270]{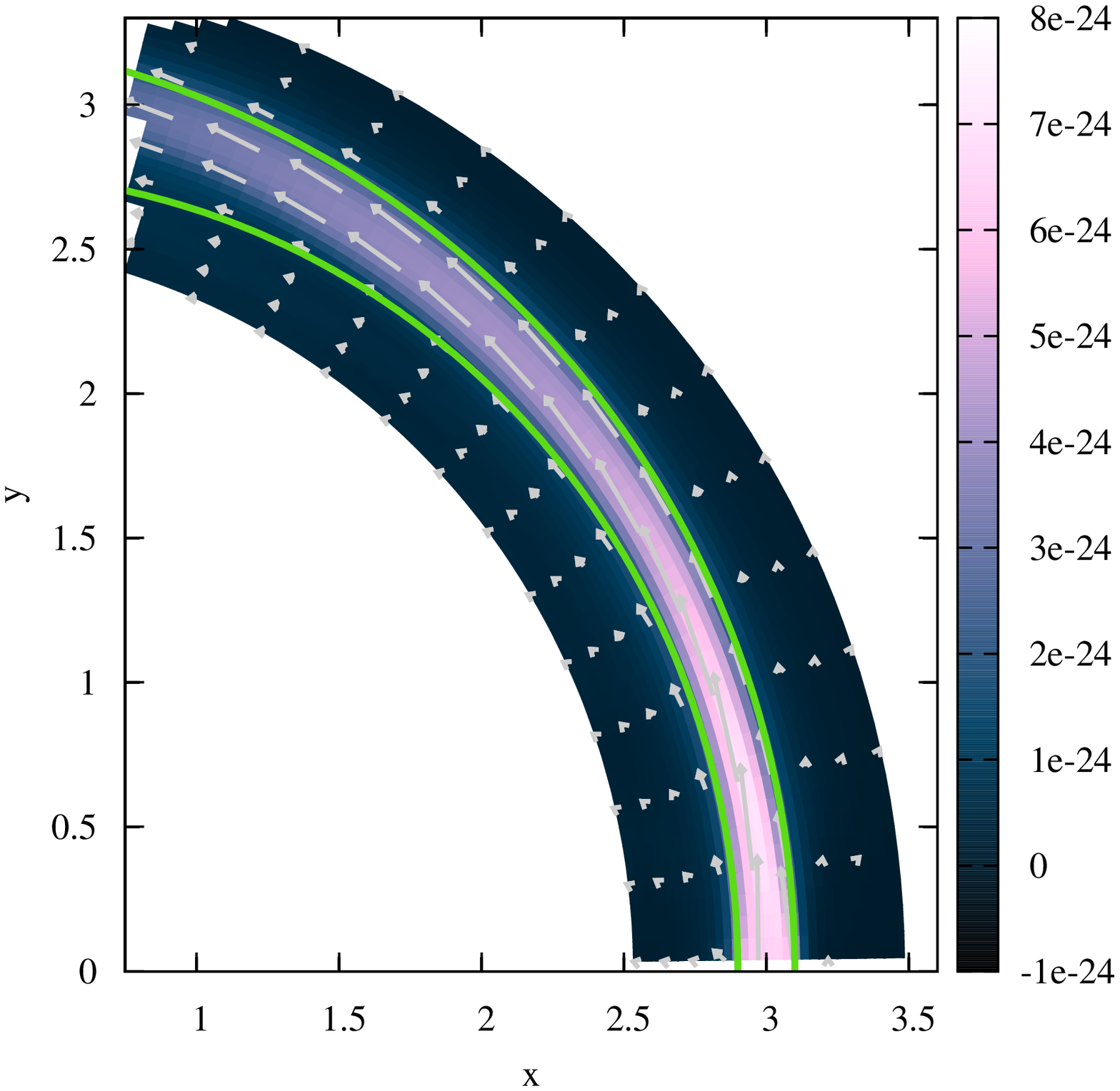}}\hspace{-.1in}
\subfigure{\includegraphics[width=.35\textwidth,angle=270]{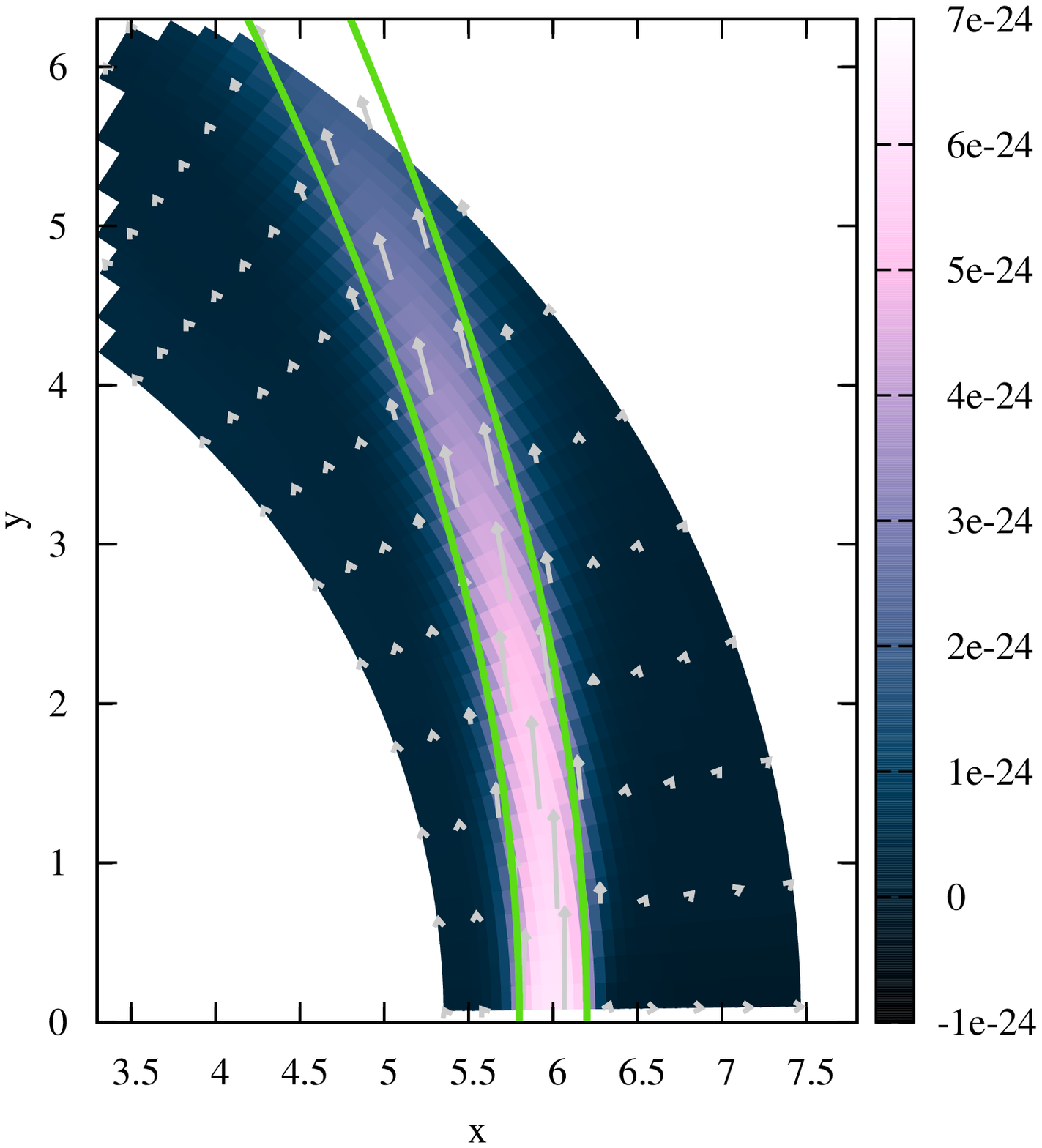}}\hspace{-.2in}
\subfigure{\includegraphics[width=.35\textwidth,angle=270]{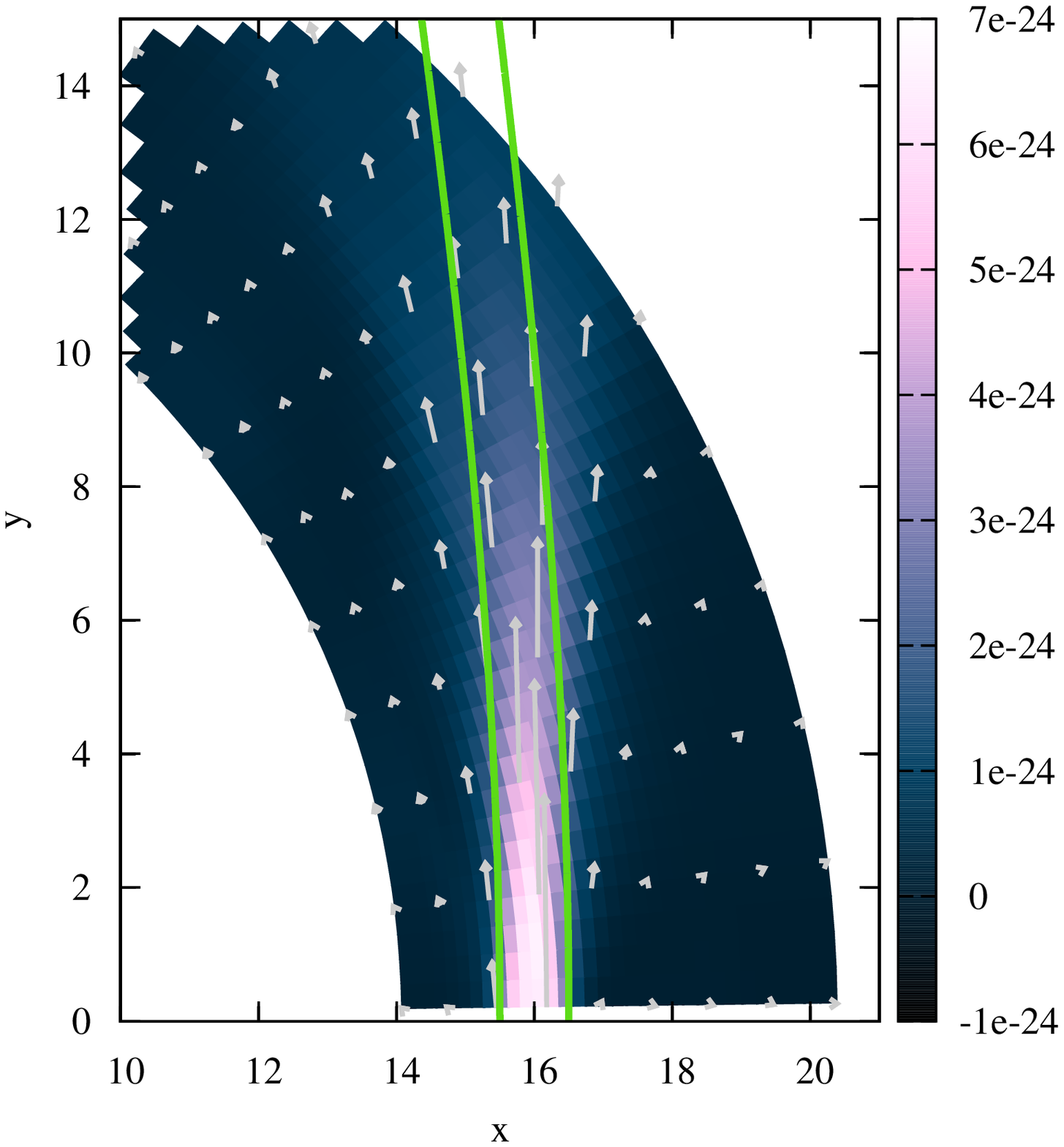}}
\caption{Results for Model 1 (left panel), Model 2 (middle), Model 3
  (right), involving light beams propagating near a Schwarzschild BH
  (see Table~\ref{t.beam} for model details). The BH is at $r=0$
  (i.e., $x=y=0$). The beams are introduced via a boundary condition
  on the $x$-axis. The beams initially move vertically, i.e., in the
  azimuthal direction. Color indicates the radiation energy density
  and arrows show the radiative flux as measured by a ZAMO. The solid
  green lines indicate true geodescis of photons at the beam
  boundaries. They were calculated using the ray-tracing code
  \texttt{GYOTO} \citep{vincentetal11}. }
  \label{f.radbeam}
\end{figure*}

The panels in Figure~\ref{f.radbeam} show the results for the
three models and geodesics of photons at beam boundaries. Consider the right panel, which corresponds to Model 3
(Table~\ref{t.beam}) with the beam centered at $r_{\rm beam}=16$. At
such a large radius we do not expect significant bending of photon
geodesics and this is indeed the case --- the beam is only slightly
bent towards the BH. We also expect the beam to be tighly confined,
i.e., it should propagate with a nearly constant width, as
indicated by the two solid green lines, the true geodesics of
photons at the boundaries of the beam. However, the numerical
solution shows significant artificial broadening. This is caused by
numerical diffusion which is significant whenever the radiative flux
vector is not aligned with the grid geometry, i.e., the beam is tilted
with respect to the grid axes.

The middle panel in Figure~\ref{f.radbeam} shows Model 2, where the
beam is centered at the marginally stable orbit: $r_{\rm beam}=6$. At
this radius, photon geodesics are significantly deviated by gravity,
resulting in strong curvature in the beam. The numerical beam follows
the correct trajectory. Moreover, numerical diffusion is lower in this
case because the curvature brings the beam into closer alignment with
the grid. There is in addition some real beam divergence because the
geodesics at different radii within the beam have different curvatures
(see the solid green lines), but this effect is not very significant.

Finally, the left panel in Figure~\ref{f.radbeam} shows Model 1, where
the center of the beam is exactly at the photon orbit: $r_{\rm
  beam}=3$.  An azimuthally oriented ray at this radius is expected to
orbit around the BH at a constant $r$. This is seen clearly in the
numerical solution.  Moreover, since the photon geodesic follows the
grid, there is practically no numerical diffusion. Indeed, there is
less diffusion than there should be (compare the numerical beam with
the two solid green lines). The beam should have some divergence as it
propagates around the BH because photons emitted inside $r=3$ curve
inward and will ultimately fall into the BH, while those emitted
outside $r=3$ curve outward and will move towards infinity. The simulated
beam does not reproduce this physical broadening very well.

\subsection{Radiative spherical accretion}

Our last test problem considers radiative spherical accretion onto a
non-rotating BH. This problem has been studied in the past by
\cite{vitello84} and \cite{nobilietal91} and more recently by
\cite{roedigetal12} and \cite{fragileetal12}. We follow
\cite{fragileetal12} in the setup of our simulations to facilitate
comparison with their results.  As in their work, we consider Thomson
scattering and thermal bremsstrahlung, which give the following
opacity coefficients,
\bea
\hspace{.7in}\kappa&=&1.7\times10^{-25} T^{-7/2} m_p^{-2} \rho\, {\rm cm^{-1}},\\
\hspace{.7in}\chi&=&\kappa+0.4 \rho\, {\rm cm^{-1}},
\eea where $\rho$ is in ${\rm g\,cm^{-3}}$ and $m_p$ is the mass of
the proton. Our numerical grid spans from $r_{\rm in}=2.5$ to $r_{\rm
  out}=2\times 10^4$ and is resolved by 512 grid points spaced
logarithmically following
$x = \log\left((r-2.2)/2\right)$ where the auxiliary variable $x$ is spaced
linearly between values corresponding to $r_{\rm in}$ and $r_{\rm out}$.
We assume a BH mass of $3 M_\odot$.
For the initial state, we choose the mass accretion rate $\dot M$
(see Table~\ref{t.bondi} for values) and set the density profile accordingly,
\be\label{eq.bondi1} \rho=-\frac{\dot M}{4\pi r^2 u^r}, \ee where the
radial velocity $u^r$ is equal to its free fall value
$u^r=-\sqrt{2/r}$. The gas temperature is given by
\be\label{eq.bondi2} T=T_{\rm out}\left(\frac\rho{\rho_{\rm
    out}}\right)^{\Gamma-1}, \ee where $T_{\rm out}$ is the
temperature at the outer radius and $\Gamma$ is the adiabatic
index. The latter is calculated from the radiation to gas pressure ratio $f_p=p_{\rm rad}/p_{\rm gas}$ of the initial state (Table~\ref{t.bondi}),
\be \Gamma=1+\frac13\left(\frac{2+2f_{\rm
    p}}{1+2f_{\rm p}}\right).  \ee The radiative energy density is set to
$E=3 f_{\rm p} p_{\rm gas}$.

The numerical simulations are run in one (radial) dimension with the
MP5 reconstruction scheme, and M1 closure for the radiation. The
primitive quantities at the outer boundary are fixed at their initial
values, as described above. At the inner boundary we apply outflow
boundary conditions, with the radial velocity fixed at the free-fall
value and rest mass density and internal energy extrapolated
proportional to $r^{-3/2}$ and $r^{-3/2\Gamma}$, respectively.
Table~\ref{t.bondi} lists the parameter values we used corresponding
to five models.  The first model, E1T6, is characterized by the lowest
mass accretion rate and is designed to highlight the ability of our
scheme to handle optically thin media. The other four models are
identical to simulations described in \cite{fragileetal12}.

\begin{table}
\caption{Models parameters for radiative spherical accretion tests}
\label{t.bondi}
\centering\begin{tabular}{@{}lcccc}
\hline
 Model & $\dot M/\dot M_{\rm Edd}$ & $T_0 [K]$ & $f_{\rm p}=\frac{p_{\rm rad}}{p_{gas}}$&$L/L_{\rm Edd}$\\
\hline
E1T6 & 1.0 & $10^6$&$1.2\times10^{-4}$ & $8.73\times10^{-8}$ \\
E10T5 & 10.0 & $10^5$&$1.2\times10^{-7}$& $3.26\times10^{-6}$ \\
E10T6 & 10.0 & $10^6$&$1.2\times10^{-4}$& $6.51\times10^{-6}$ \\
E10T7 & 10.0 & $10^7$&$1.2\times10^{-1}$& $1.45\times10^{-5}$ \\
E100T6 & 100.0 & $10^6$&$1.2\times10^{-4}$& $2.00\times10^{-4}$ \\
\hline
\end{tabular}\\
Model names and parameters after \cite{fragileetal12}.
 \end{table}

\begin{figure*}
  \centering
\subfigure{\includegraphics[width=1.3\columnwidth,angle=270]{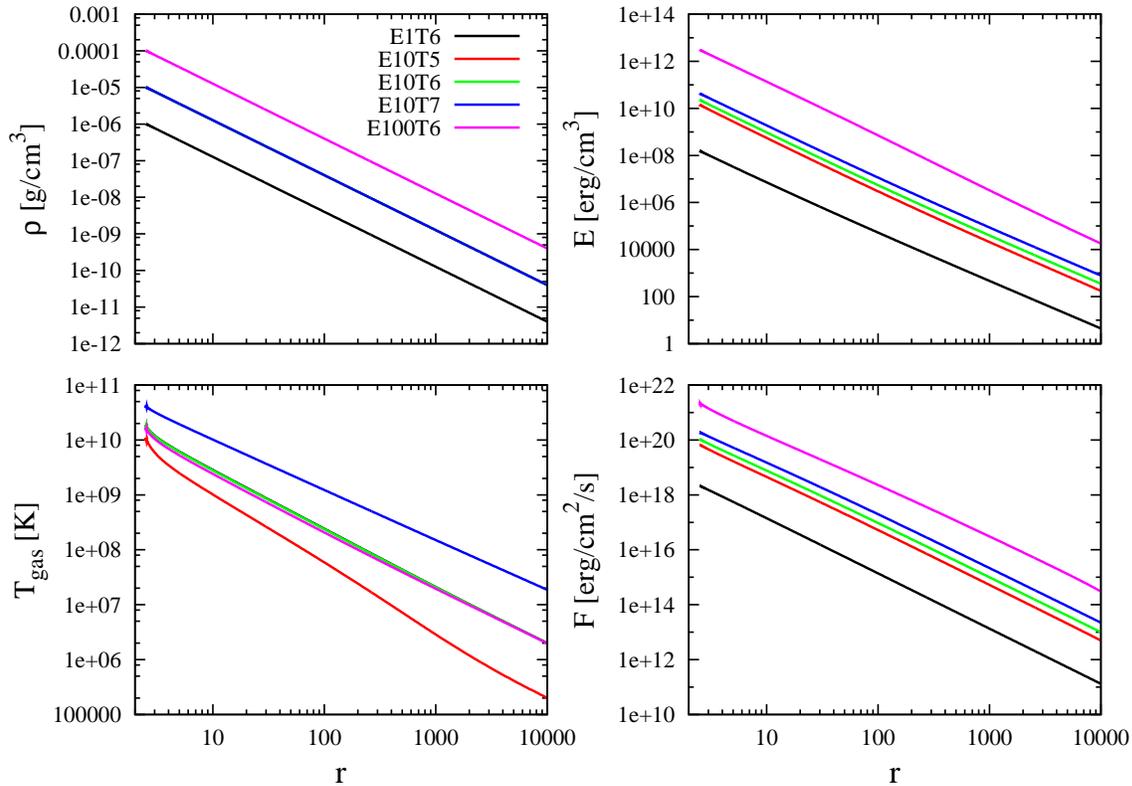}}
\caption{Numerical results obtained with \koral\ for five models of
  spherical Bondi accretion with radiation. The panels show density
  (top-left), radiative energy density (top-right), gas temperature
  (bottom-left) and radiative radial flux (bottom-right).  Parameters
  of the models are given in Table~\ref{t.bondi}. The results show
  that the code handles optically-thin and optically-thick regions
  equally well, without producing unphysical oscillations.}
  \label{f.bondi}
\end{figure*}

Figure~\ref{f.bondi} shows the numerical solutions obtained with
\koral. The top-left panel presents profiles of density, which follow
the initial profile (equation~\ref{eq.bondi1}) throughout the
simulation. The bottom-left panel shows the gas temperature. For all
but the coldest model, E10T5, the temperature follows
equation~(\ref{eq.bondi2}). In the case of model E10T5, the gas is
hotter than the analytical result.  This is because of gas-radiation
coupling which heats up the gas as it approaches the BH (the
analytical solution assumes that there is no interaction). The small
kinks in the temperature profiles near the inner boundary are an
artifact of the inner boundary condition. They do not influence the
rest of the solution since information cannot travel upstream in the
supersonic flow.

The top-right and bottom-right panels in Figure~\ref{f.bondi} show
radial profiles of the radiative energy density and flux for the five
models.  Both quantities follow roughly an $r^{-2}$ scaling,
reflecting the fact that in steady-state (barring redshift factors)
the luminosity is equal to $4\pi Fr^{2}$ and should be conserved.

Because the flux in these models is non-negligible compared to the
energy density (e.g., $F\approx 0.9E$ for the E10 family of models),
the Eddington closure scheme does not work very well, especially at
low optical depth.  For instance, \cite{fragileetal12} used Eddington
closure and obtained unphysical noise or breaks in their profiles of
radiative quantities (see their Figure~5) in all models with $\dot M <
300 \dot M_{\rm Edd}$. This just reflects the fact that their closure
cannot handle optically thin media.  Our algorithm uses the M1 closure
scheme and has no problems with either optically thick or thin
regimes. To emphasize this point, we have solved an additional model,
E1T6, in which the accretion rate is an order of magnitude lower than
the smallest rate considered by \cite{fragileetal12}. \koral\ works
fine for this model, and can, in fact, handle even more extreme
configurations, both at lower and higher accretion rates.

\begin{figure}
  \centering
  \subfigure{\includegraphics[width=.7763\columnwidth,angle=270]{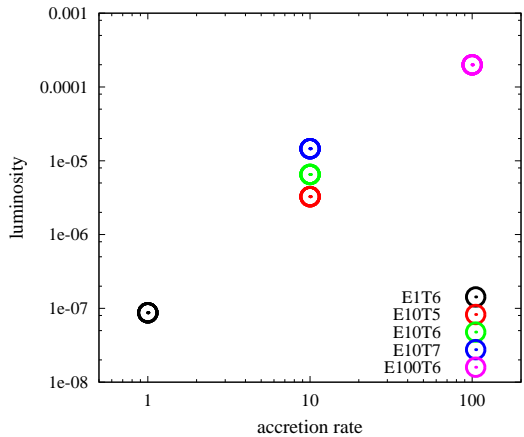}}
  \caption{Luminosity (in Eddington units) emerging from a spherically
    symmetric radiative Bondi accretion flow onto a non-rotating BH,
    plotted as a function of the dimensionless (reduced) accretion
    rate in units of the Eddington accretion rate, $\dot M/\dot M_{\rm
      Edd}$.  Parameters of the models are given in
    Table~\ref{t.bondi}. Low accretion rates, where the accretion flow
    becomes optically thin, are usually problematic for numerical
    codes. \koral\ has no problem handling either low or high
    accretion rates. }
  \label{f.bondi2}
\end{figure}

For direct comparison of our results with those reported in
\citet{fragileetal12}, we have calculated for all our models the
luminosities,
\be 
L=4\pi Fr^2, 
\ee
emerging at radius $r=10^3$. Our results, shown in
Figure~\ref{f.bondi2}, are consistent with those obtained by
\cite{fragileetal12} (compare their Figure~6). Note, however, an
important difference. The scheme used by \citet{fragileetal12} is
explicit, therefore it cannot reliably treat optically-thin
media. This causes the luminosities they quote to be sensitive to the
radius at which they are measured. For instance, reading off the
luminosity values in their simulations (Figure~5 of their paper) at a
distance at which the flow is optically thin, e.g., $r = 10^4$, one
obtains erroneous lower values of the luminosity.

\section{Summary}
\label{s.summary}

In this paper we have introduced a semi-implicit numerical scheme for
general relativistic radiation hydrodynamics.  The scheme is based on
a covariant formulation of the M1 closure scheme for the radiation
moments. The radiative source terms are handled semi-implicitly, and
hence this approach can handle practically all optical depths.  The
algorithm has been implemented and tested in a new GRRHD code \koral.
It can be easily incorporated into any general relativistic
hydrodynamic or magnetohydrodynamic conservative code.

Our tests indicate that {\koral} works well for a variety of physical
regimes and geometries: optically thick vs optically thin, gas
dominated vs radiation dominated, flat space vs curved space. Also, as
expected, we find that M1 closure has some advantages over the
standard Eddington closure scheme in the case of optically-thin media:
namely, it accurately propagates light rays and it is able to resolve
shadows.

The semi-implicit radiation scheme implemented in {\koral} does not
overwhelmingly slow down the code. Apart from the fact that the
inclusion of radiation introduces 4 new conserved quantities compared
to pure hydrodynamics, the only important computational steps that
affect performance are (i) calculation of the radiative characteristic
wavespeeds (Section~\ref{s.wavespeeds}), and (ii) numerical solution
of the system of four non-linear equations that arise in the implicit
treatment of radiative source terms (Section~\ref{s.radsource}).  We
expect each of these steps could be speeded up with more effort.
However, even without further improvements, the code performance is
already sufficiently good that global GRRHD or GRRMHD simulations of
accretion disks with near- or super-Eddington luminosities appear
feasible.  Meanwhile, it would be of great interest to develop closure
schemes beyond M1, e.g., by directly evolving the photon
distribution function and using it to close the radiation moments, for conservative GR codes.

\section{Acknowledgements}

The authors thank Frederic Vincent for calculating the photon
trajectories shown in Fig.~\ref{f.radbeam}, James M.~Stone for
insightful discussions, and the referee for helpful comments. RN and AS were supported in part by NASA grant
NNX11AE16G. AT was supported by the Princeton Center for
Theoretical Science Fellowship.  We acknowledge NSF support via XSEDE
resources, NICS Kraken and Nautilus under grant numbers TG-AST080026N
(RN and AS) and TG-AST100040 (AT), and NASA support via High-End
Computing (HEC) Program through the NASA Advanced Supercomputing (NAS)
Division at Ames Research Center (AS and RN) that provided access to
the Pleiades supercomputer.
 
\bibliographystyle{mn2e}

\appendix

\section{Explicit-implicit method for the radiative source operator}
\label{ap.expimp}
In Sect.~\ref{s.radsource} we described a semi-implicit method for
applying the radiative source terms $G^\mu$ in the lab frame. It
requires solving a four-dimensional system of non-linear
equations. Because of that fact, the numerical efficiency is limited.
Furthermore, the method may sometimes fail to produce a solution,
e.g., when the initial guess is not close enough to the true solution
or when the solution at an intermediate iteration becomes unphysical
such as having $\widehat F>c\widehat E$. To handle these situations we
have developed the approximate analytical method described in this
Appendix, which is both robust and failsafe. We use it as a backup
solver for the fiducial algorithm. Its limitations are discussed
below.

Let us assume that the advective and
geometric source terms have already been applied as per steps
(i)--(ix) in Section~\ref{s.algorithm}. The only remaining terms are
the radiative forces. For instance, equations~(\ref{eq.radlab1}) require us
to time-evolve \be\label{eq.radsource} \partial_t( R^t_\nu)=-G_\nu.
\ee Let us assume to start with that we treat this term explicitly.
Then, the update of the conserved quantities is given simply by
\be\label{eq.impl0} \Delta R^t_\nu=-G_\nu \Delta t.  \ee The right
hand side can be rewritten as \be\label{eq.impl1a} -G_\nu \Delta t = -
g_{\mu\nu} G^{\mu} \Delta t = -g_{\mu\nu} \Lambda^\mu_\alpha \widehat
G^\alpha \Delta t, \ee where $\widehat G^\alpha$ is the radiative
four-force in the fluid frame.

In the same explicit spirit, the fluid-frame source terms in equations~(\ref{eq.radff}) may be
written as
\bea\label{eq.impl2b}
\hspace{1.2in}\Delta \widehat E& =& -\widehat G^t \Delta t,\\ \nonumber
\hspace{1.2in}\Delta \widehat F^j&=& -\widehat G^j \Delta t.  \eea
 These updates of $\widehat E$ and $\widehat F^j$ generally do not
correspond to the updates of the conserved quantities we search for
because the spatial and temporal derivatives are mixed when moving
from one frame to another, i.e., the operator splitting is
frame-dependent. However, comparing the right hand sides of
equations~(\ref{eq.impl1a}) and (\ref{eq.impl2b}), it is clear that the change
in the conserved quantities is related to the updates of the
primitive radiative quantities $\widehat E$ and $\widehat F^i$ calculated purely in the
fluid frame. From equations~(\ref{eq.impl0}) and (\ref{eq.impl1a}) it follows that
\be \Delta R^t_\nu = g_{\mu\nu} \Lambda^\mu_\alpha
\widehat\Delta_P^\alpha, \ee where components of the
four-vector of updates of primitive radiative quantities in the
fluid frame, $\widehat\Delta_P$ (calculated according to
equation~\ref{eq.impl2b}), are given by the following 4-vector, \be
\widehat\Delta_P = [\Delta \widehat E, \Delta \widehat F^j]. \ee 
This quantity may be found using either
explicit or implicit methods.

The explicit approach is the simplest and fastest in the
optically thin regime, but it fails in the
optically thick regime. To calculate $\widehat \Delta_P$ via the 
explicit approach, we simply set
\be
\widehat\Delta_P = [-\widehat G^t_{(n)} \Delta t, -\widehat G^j_{(n)} \Delta t],
\ee
where the subscript $(n)$ indicates variables evaluated at the
beginning of the current time step.

The implicit approach works well for all optical depths, and is
especially important at high optical depths, but it usually
involves more computations. In principle, one should solve the
following set of fluid-frame equations for gas internal energy, momenta, radiative
energy and fluxes,
\bea
\label{eq.impl1}\hspace{.4in}u_{(n+1)}-u_{(n)} &=& \kappa(\widehat E_{(n+1)}-4\sigma T^4_{(n+1)}) \Delta t,\\
\label{eq.impl2}\hspace{.4in}m^{j}_{(n+1)}-m^{j}_{(n)} &=& \chi \widehat F^{j}_{(n+1)} \Delta t,\\
\label{eq.impl3}\hspace{.4in}\widehat E_{(n+1)}-\widehat E_{(n)} &=& - \kappa(\widehat E_{(n+1)}-4\sigma T^4_{(n+1)}),\\
\label{eq.impl4}\hspace{.4in}\widehat F^{j}_{(n+1)}-\widehat F^{j}_{(n)} &=& - \chi \widehat F^{j}_{(n+1)} \Delta t,
\eea
where the subscripts $(n)$ and $(n+1)$ denote values at the beginning and the end of the
current time step, respectively, and $\kappa$ and $\chi$
are computed from the gas properties at time $(n+1)$. As a simplification,
we assume that $\kappa$ and $\chi$ do not change significantly during a single time step, and we set
\bea
\hspace{1.2in}\kappa &=& \kappa(\rho,T_{(n)}),\\
\hspace{1.2in}\chi   &=& \chi(\rho,T_{(n)}).
\eea
Now, equation~(\ref{eq.impl4}) becomes decoupled from the others and may be solved directly for the updated fluxes,
\be
\widehat F^j_{(n+1)}=\frac{\widehat F^j_{(n)}}{1+\chi \Delta t}.
\ee

To find the new value of the radiative energy density one has to solve
equations~(\ref{eq.impl1}) and (\ref{eq.impl3}) together. Taking into account the ideal gas equation of state,
\be
p=(\Gamma-1)u=\frac{\cal R}{\mu}\rho T,
\ee
where $\mu$ is the mean molecular weight, it is straightforward to
combine these two equations into one quartic equation for 
$\widehat E_{(n+1)}$. {\koral}
solves this equation numerically using the Newton method with
$\widehat E_{(n)}$ as the initial guess. However, 
analytical solvers for quartic
equations may also be implemented; in fact, a linearized version of the
quartic is often quite adequate.
Once $\widehat{E}$ and $\widehat{F}^j$ have been calculated, the four-vector of updates, $\widehat{\Delta}_P$, is given by,
\be
\label{eq.delta.impl}
\widehat \Delta_P =[\widehat E_{(n+1)}-\widehat E_{(n)},\widehat F^j_{(n+1)}-\widehat F^j_{(n)}].
\ee
This update is applied in step (xi) of the algorithm (see Section~\ref{s.algorithm}) in case the fiducial lab frame implicit method fails.

The updates of the conserved quantities calculated using this approach
diverge from the fiducial semi-analytical results in the limit of
large velocites $u^t\gg1$ and large optical depths per cell
$\tau\gg1$. Therefore, in general, one should use this algorithm only
as a failsafe backup method. However, if one is confident that the
problem at hand does not involve such conditions, one might consider
using this method as the default, thereby increasing the code
efficiency.

\end{document}